\documentclass[aps,prl,showpacs,twocolumn,superscriptaddress,nobibnotes]{revtex4-1}

\RequirePackage{subfigure}
\usepackage{lipsum}
\usepackage{graphicx}
\usepackage{color}
\usepackage{enumerate}
\usepackage{amsmath}
\usepackage{bm}
\usepackage{diagbox}
\usepackage{amssymb}
\usepackage{stmaryrd}
\usepackage{multirow}
\usepackage{bbm}
\usepackage[normalem]{ulem}
\usepackage{float}
\usepackage{longtable,booktabs}
\usepackage[dvipsnames,table,xcdraw]{xcolor}
\usepackage[normalem]{ulem}
\usepackage{physics}
\usepackage{url}
\usepackage{soul} 
\usepackage[colorinlistoftodos]{todonotes}
\usepackage{verbatim}
\usepackage{graphicx}
\usepackage{subfigure}
\graphicspath{{figures/}}
\usepackage{appendix}
\definecolor{darkblue}{rgb}{0.1,0.2,0.6} \definecolor{darkred}{rgb}{0.8,0.1,0.2}
\usepackage[colorlinks,citecolor=darkblue,linkcolor=darkred,urlcolor=darkblue]{hyperref}
\usepackage[all]{hypcap} 
\usepackage[draft]{pdfcomment}

\newcommand{\cmmnt}[1]{}

\def\beq{\begin{equation}}
\def\eeq{\end{equation}}

\begin{document}
\long\def\/*#1*/{}

\title{Autoregressive Transformer Neural Network for Simulating Open Quantum Systems via a Probabilistic Formulation}

\author{Di Luo}
\thanks{Co-first authors.}
\affiliation{Department of Physics,  University of Illinois at Urbana-Champaign, IL 61801, USA}
\affiliation{IQUIST and Institute for Condensed Matter Theory, University of Illinois at Urbana-Champaign} 
\author{Zhuo Chen}
\thanks{Co-first authors.}
\affiliation{Department of Physics, University of Illinois at Urbana-Champaign, IL 61801, USA} 
\author{Juan Carrasquilla}
\affiliation{Vector Institute for Artificial Intelligence, MaRS Centre, Toronto, Ontario, Canada}
\affiliation{Department of Physics and Astronomy, University of Waterloo, Ontario, N2L 3G1,Canada}
\author{Bryan K. Clark}
\affiliation{Department of Physics,  University of Illinois at Urbana-Champaign, IL 61801, USA}
\affiliation{IQUIST and Institute for Condensed Matter Theory, University of Illinois at Urbana-Champaign} 
\affiliation{ NCSA Center for Artificial Intelligence Innovation,University of Illinois at Urbana-Champaign}

\begin{abstract}
The theory of open quantum systems lays the foundations for a substantial part of modern research in quantum science and engineering. Rooted in the dimensionality of their extended Hilbert spaces, the high computational complexity of simulating open quantum systems calls for the development of strategies to approximate their dynamics. 
In this paper, we present an approach for tackling open quantum system dynamics.  Using an exact probabilistic formulation of quantum physics based on positive operator-valued measure (POVM), we compactly represent quantum states with autoregressive transformer neural networks; such networks bring significant algorithmic flexibility due to efficient exact sampling and tractable density. We further introduce the concept of String States to partially restore the symmetry of the autoregressive transformer neural network and improve the description of local correlations. Efficient algorithms have been developed to simulate the dynamics of the Liouvillian superoperator using a forward-backward trapezoid method and find the steady state via a variational formulation.
Our approach is benchmarked on prototypical one and two-dimensional systems, finding results which closely track the exact solution and achieve higher accuracy than alternative approaches based on using Markov chain Monte Carlo to sample restricted Boltzmann machines.
Our work provides general methods for understanding quantum dynamics in various contexts, as well as techniques for solving high-dimensional probabilistic differential equations in classical setups.

\end{abstract}
\maketitle

\textit{Introduction.} While the universe itself is a closed quantum system, all other systems within the universe are open quantum systems coupled to the environment around them.  Open quantum systems (OQS) play a crucial role in fundamental quantum science, quantum control and quantum engineering \cite{Verstraete2009,Barreiro2011}.   In recent years, there has been a significant interest both theoretically and experimentally in better understanding open quantum systems \cite{Sieberer_2016,PhysRevB.93.014307,PhysRevA.92.022116,PhysRevLett.114.220601,Jaschke_2018,PhysRevLett.116.237201,PhysRevB.97.035103,PhysRevX.6.031011,PhysRevLett.115.080604,PhysRevB.95.134431,PhysRevLett.122.110405,PhysRevA.97.062107,PhysRevA.97.052129,PhysRevA.98.063815,PhysRevA.99.032115,Kshetrimayum2017,
RevModPhys.85.299,Hartmann_2016,Noh_2016,PhysRevA.94.033841,PhysRevA.96.023839,PhysRevA.96.043809,PhysRevX.5.031028,PhysRevA.95.012128,PhysRevA.93.033824,PhysRevX.7.011012,Fink2018,PhysRevX.7.011016,PhysRevA.95.043826,PhysRevA.86.012116,PhysRevLett.116.070407,PhysRevA.96.033826,PhysRevLett.110.195301,PhysRevA.97.013853,PhysRevX.6.031011,PhysRevLett.110.257204,Rota_2018,PhysRevB.95.134431,PhysRevA.97.062107}. In the field of quantum engineering, coupling to the environment generates decoherence driving the destruction of entanglement within quantum devices. Quantum computers rely on the qubit-environment coupling to apply quantum gates as well as try to minimize unwanted coupling to mitigate errors on the qubits \cite{alex2020circuit}.

Unlike closed quantum states which can be represented by a wavefunction, the density matrix $\rho$ becomes the core object of study in open quantum systems.  A typical model of an OQS evolves the density matrix under both the Hamiltonian $H$ as well as a series of dissipative operators which transfer energy and information out to a featureless bath leading to the Lindblad equation,
\begin{equation}
     \dot{\rho} = \mathcal{L}\rho
                \equiv -i[{H}, \rho] + \sum_k\frac{\gamma_k}{2}\left(2\Gamma_k\rho \Gamma_k^\dagger - \{\rho, \Gamma_k^\dagger \Gamma_k\}\right),
\label{equ:lind-orig}
\end{equation}
where $\gamma_k$ are the dissipation rates associated with jump operators $\Gamma_k$. Although there is hope that quantum algorithms \cite{vqe_open,dongling_open,lee2020neuralnetwork,ramusat2020quantum,liu2021variational} may eventually overcome the simulation bottlenecks in OQS, a direct solution to the Lindblad equation is difficult because the Hilbert space grows exponentially with the number of particles, making classical simulations largely intractable. To deal with this curse of dimensionality, OQS have historically been studied with renormalization group approaches \cite{PhysRevLett.115.080604,PhysRevB.95.134431,PhysRevLett.122.110405}, mean field methods \cite{PhysRevB.97.035103,PhysRevX.6.031011,scarlatella2020dynamical}; or simulated with tensor networks \cite{PhysRevA.92.022116,PhysRevLett.93.207204,PhysRevLett.93.207205,PhysRevLett.114.220601,Jaschke_2018,PhysRevLett.116.237201,Kshetrimayum2017} which compress the density matrix.  Unfortunately, while tensor networks have proved fruitful in one dimension, their use for OQS in higher dimensions has been severely limited. Recently, inspired by the advances in the description of many-body systems in terms of neural network wavefunctions \cite{lagaris1997,Carleo_2017, PhysRevLett.122.226401,ferminet,paulinet,PhysRevResearch.2.023358,PhysRevLett.124.020503,topo_wf,Gao2017,Glasser_2018}, ideas from machine learning have been applied to OQS studying real-time dynamics in one dimension (1-D), steady states in one and two dimensions (2-D) \cite{RBM_Vicentini,PhysRevB.99.214306,PhysRevLett.122.250502,PhysRevLett.122.250501} and determining the Liouvillian gap \cite{yuan2020solving} by representing the density matrix as a restricted Boltzmann machine (RBM) \cite{PhysRevLett.120.240503}.

Here, we outline an alternative approach to using machine learning ideas to simulate the Lindblad equation.  Many machine learning architectures and generative models (such as the RBM) have fundamentally been designed to represent probability distributions (e.g. probability distributions over images on the internet) making them inadequate to store quantum states, which are complex valued in general. To overcome this, novel approaches have been devised such as using complex weights within RBM; despite these innovative ideas, effectively representing states with signs have been a key bottleneck in this field \cite{PhysRevB.100.125131,westerhout2020,PhysRevResearch.2.033075}.  

This motivation has inspired us to utilize the recent developments in the probabilistic formulation of quantum mechanics \cite{Lundeen2009,2019rec,quantum_circuit,Kiktenko_2020} to simulate the Lindblad equation.  In this formulation the state is mapped to a probability distribution which we represent compactly using the Transformer \cite{transformer}---a machine learning architecture from which one can efficiently sample the probability distribution exactly.  Using this, we develop efficient algorithms to both update the state of the Transformer under dynamic evolution as well as find the Transformer which represents the steady state of the Lindblad equation.  
To perform the dynamic evolution, we combine the second-order forward-backward trapezoid method \cite{iserles_2008} with stochastic optimization on the Transformer. 
Since the Transformer does not naively preserve the symmetry of the true dynamic (or fixed-point) state, we further improve upon our results by developing an additional ansatz---string states---which explicitly restores some of these symmetries.  We proceed to benchmark this work on a series of one- and two-dimensional systems.

\textit{Lindblad Equation as a Probability Equation. }
The general objective of this paper is to develop an approach to solve for the dynamics and fixed point of the density matrix $\rho$ in the Lindblad equation  (Eq.~\ref{equ:lind-orig}). 
We test this approach on
 two models---the transverse-field Ising model (TFIM), where 
$H = \frac{V}{4}\sum_{\langle i, j\rangle}\sigma_i^{(z)}\sigma_j^{(z)} + \frac{g}{2}\sum_k\sigma_k^{(x)},
$
and the Heisenberg model, where 
$H = \sum_{\langle i, j\rangle}\sum_{w=x,y,z}J_w\sigma_i^{(w)}\sigma_j^{(w)} + B\sum_k \sigma_k^{(z)}.
$
In both cases, $\Gamma_k = \sigma_k^{(-)} = \frac{1}{2}(\sigma_k^{(x)} - i\sigma_k^{(y)})$.
We are interested in the expectation values of local observables given by the Pauli matrices averaged over all qubits, i.e. for a system with $n$ qubits, we consider $\langle \sigma_w \rangle = \frac{1}{n}\sum_{i}\langle\sigma_i^{(w)}\rangle$ for $w = x, y, z$.
Typically, the density matrix $\rho$ is represented (explicitly or implicitly) in an orthogonal basis. In this work, we instead represent $\rho$ in the POVM formalism. Given an informationally complete POVM (IC-POVM), a density matrix $\rho$ of a spin-$1/2$ system can be uniquely mapped to a probability distribution $p(\bm{a})$, where $\bm{a}$ spans over all $4^n$ measurement outcomes in the POVM basis. An IC-POVM is defined by a collection of positive semi-definite operators $\{M_{(\bm{a})}\}$ called the frame, which specifies the probability distribution $p(\bm{a}) = \Tr(\rho M_{(\bm{a})})$.
The inverse transformation is given by $\rho = \sum_{\bm{b}} p(\bm{b})N^{(\bm{b})}$,
where the dual-frame $\{N^{(\bm{b})}\}$ can be computed 
from the frame as $N^{(\bm{b})} = \sum_a M_{(\bm{a})}T^{-1}_{\bm{a}\bm{b}}$. The elements of the overlap matrix $T$ are given by  $T_{\bm{a}\bm{b}}= \Tr(M_{(\bm{a})} M_{(\bm{b})})$, and $T^{-1}_{\bm{a}\bm{b}}$ represent the elements of the inverse overlap matrix $T^{-1}$.
Thus, we can re-express the Lindblad equation as
\begin{equation}
    \dot{p}(\bm{a}) = \sum_{\bm{b}} p(\bm{b}) L_{\bm{a}}^{\bm{b}}
              = \sum_{\bm{b}} p(\bm{b}) \left(A_{\bm{a}}^{\bm{b}} + B_{\bm{a}}^{\bm{b}}\right),
\label{equ:lind-povm}
\end{equation}
with 
\begin{align}
    A_{\bm{a}}^{\bm{b}} &= -i\Tr\left({H}[{N}^{(\bm{b})}, {M}_{(\bm{a})}]\right); \nonumber \\
    B_{\bm{a}}^{\bm{b}} &= \sum_k\frac{\gamma_k}{2}\Tr\left(2\Gamma_k{N}^{(\bm{b})}\Gamma_k^\dagger{M}_{(\bm{a})} - \Gamma_k^\dagger\Gamma_k\{{{N}^{(\bm{b})}}, {{M}_{(\bm{a})}}\}\right).
\end{align}
We work with an IC-POVM where the frame and dual-frame
are constructed from local frames acting on single spins as $\{M_{(\bm{a})}\} = \{M_{(a_1)} \otimes M_{(a_2)} \otimes M_{(a_3)} \otimes \cdots\}$ and $\{N^{(\bm{b})}\} = \{N^{(b_1)} \otimes N^{(b_2)} \otimes N^{(b_3)} \otimes \cdots\}$
with four outcomes per spin $a_i$.  This allows us to write $p(\bm{a}) = p(a_1, a_2, a_3, \cdots)$.
The expectation value of observables are given by  
$
    \langle O \rangle = \sum_{\bm{b}}p(\bm{b})\Tr\left(ON^{\bm{b}}\right) \approx \frac{1}{N_s}\sum_{\bm{b}\sim p}^{N_s}\Tr\left(ON^{\bm{b}}\right),
$
where $N_s$ is the number of samples $\bm{b}$ drawn from the distribution $p(\bm{b})$ used to estimate $\langle O \rangle$. We emphasize that a complete specification of the probability distribution $p(\bm{b})$ requires $4^n$ probability values for an $n$-site system.  

\textit{Autoregressive Models and String States.}
We have chosen to model the probability distribution in a compact way with an autoregressive neural network where the probability of a given configuration $\bm{a}$ is expressed through its conditional probabilities $p_\theta (\bm{a}) = \prod_k p_\theta(a_k| a_1, a_2, \cdots, a_{k-1})$.
This representation allows for exact sampling of a configuration from the space of probability distributions without invoking Markov chain Monte Carlo techniques. 
Modern incarnations of autoregressive models include, among others,  recurrent neural networks (RNN) \cite{lstm,gru}, pixel convolutional neural networks (PixelCNN) \cite{pixelcnn}, Transformers \cite{transformer}. Recent work has effectively applied these models to quantum systems \cite{2019rec,PhysRevResearch.2.023358,PhysRevLett.124.020503,quantum_circuit,cha2020}.
Here, we use an autoregressive Transformer, which follows the same architecture as the model in \cite{quantum_circuit}.
The Transformer consists of two hyper-parameters: the number of transformer layers stacked on each other $n_l$ and the hidden dimension $n_d$, which we adjusted for different tests. 

\begin{figure}[ht]
    \centering
    \subfigure[ String 0.]{
    \includegraphics[width=0.16\linewidth]{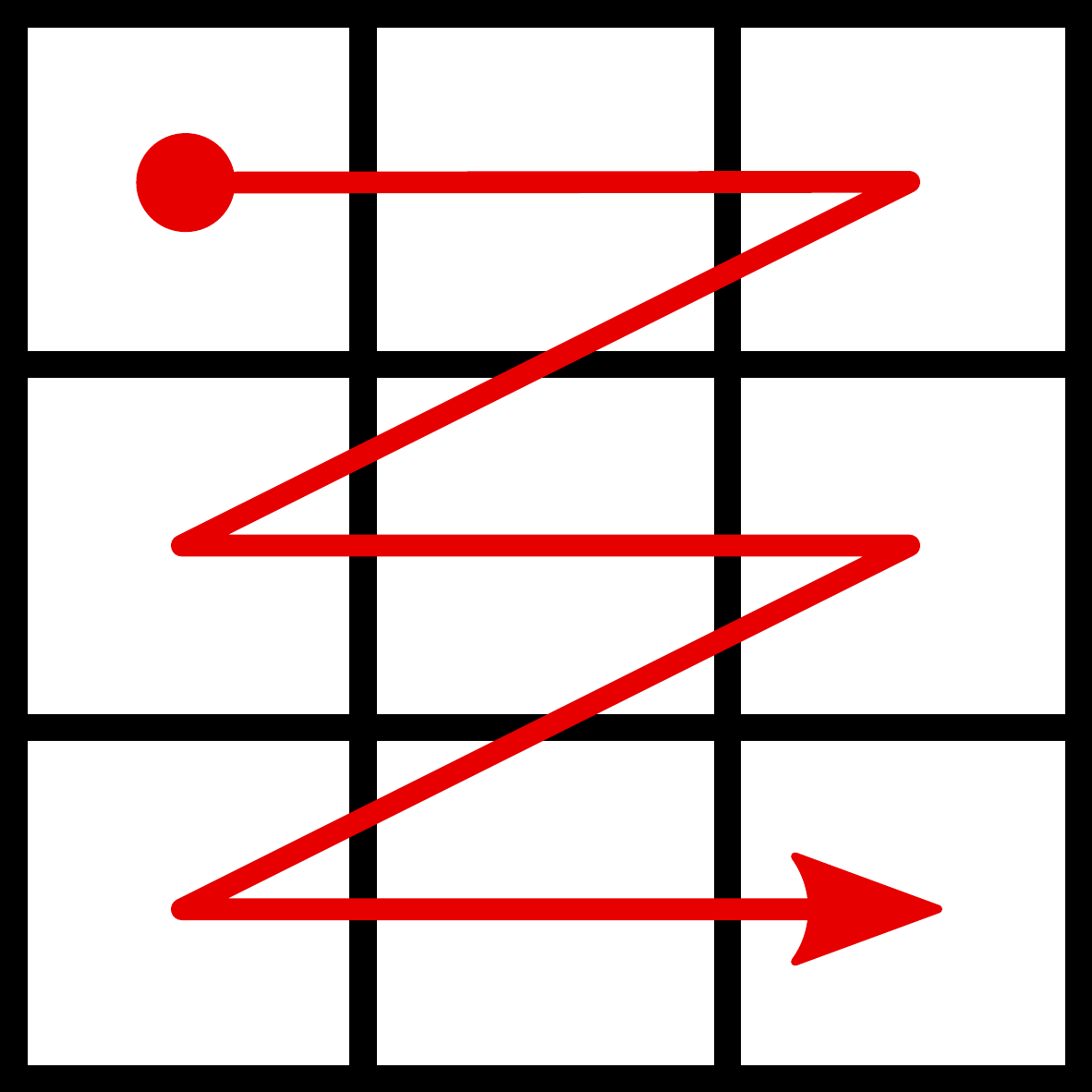}
    \label{fig:string0}}
    \subfigure[String 1.]{
    \includegraphics[width=0.16\linewidth]{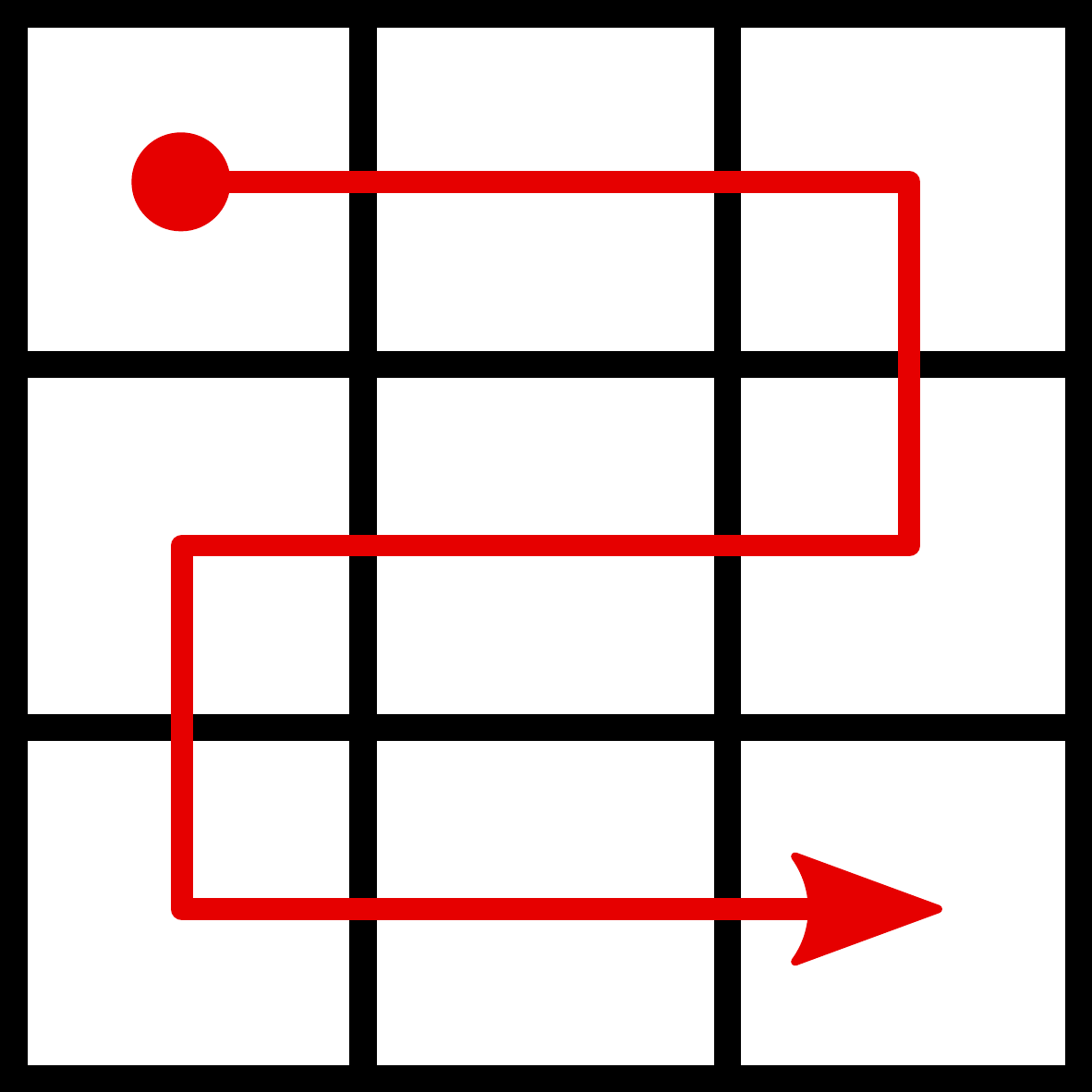}
    \label{fig:string1}}
    \subfigure[String 2.]{
    \includegraphics[width=0.16\linewidth]{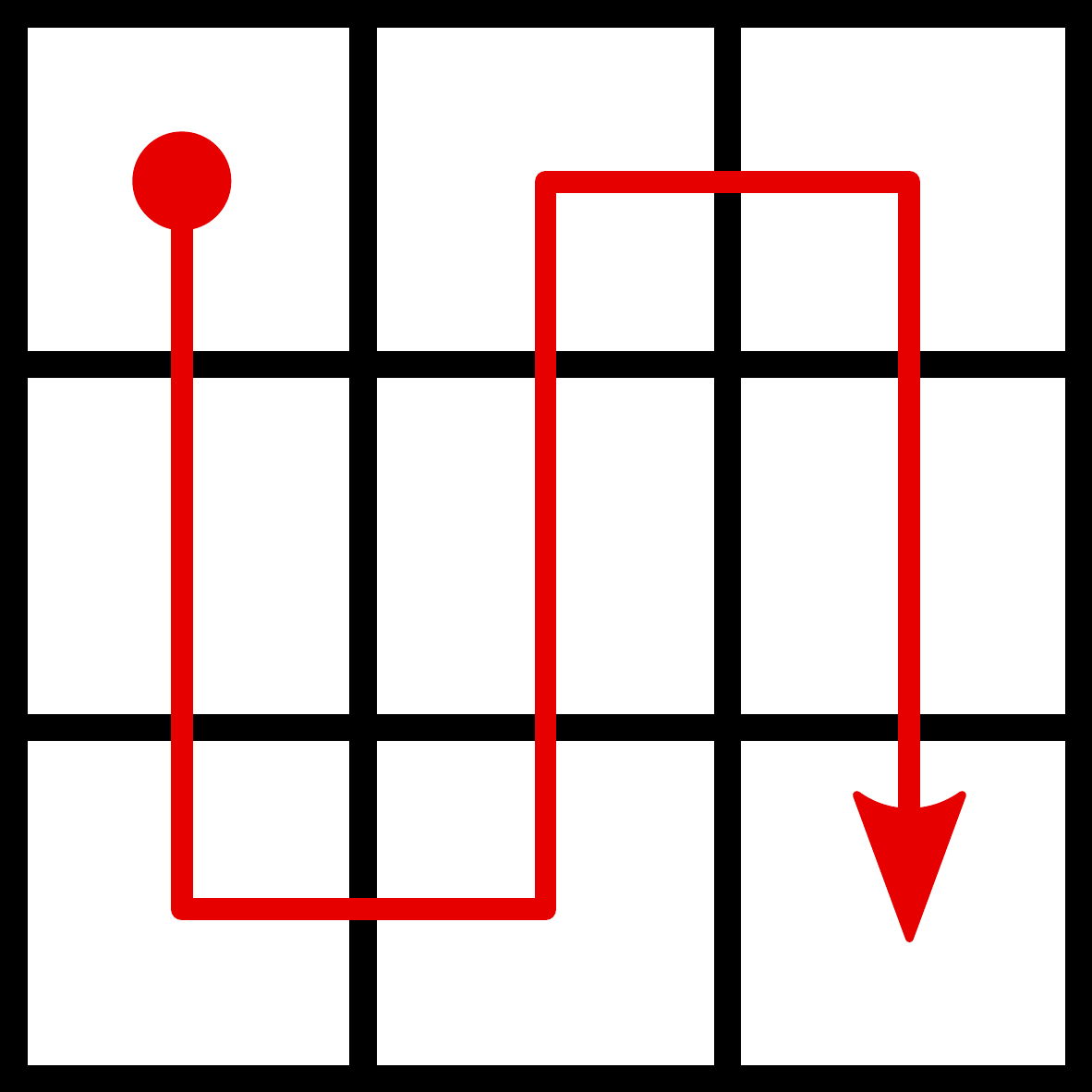}
    \label{fig:string2}}
    \subfigure[String 3.]{
    \includegraphics[width=0.16\linewidth]{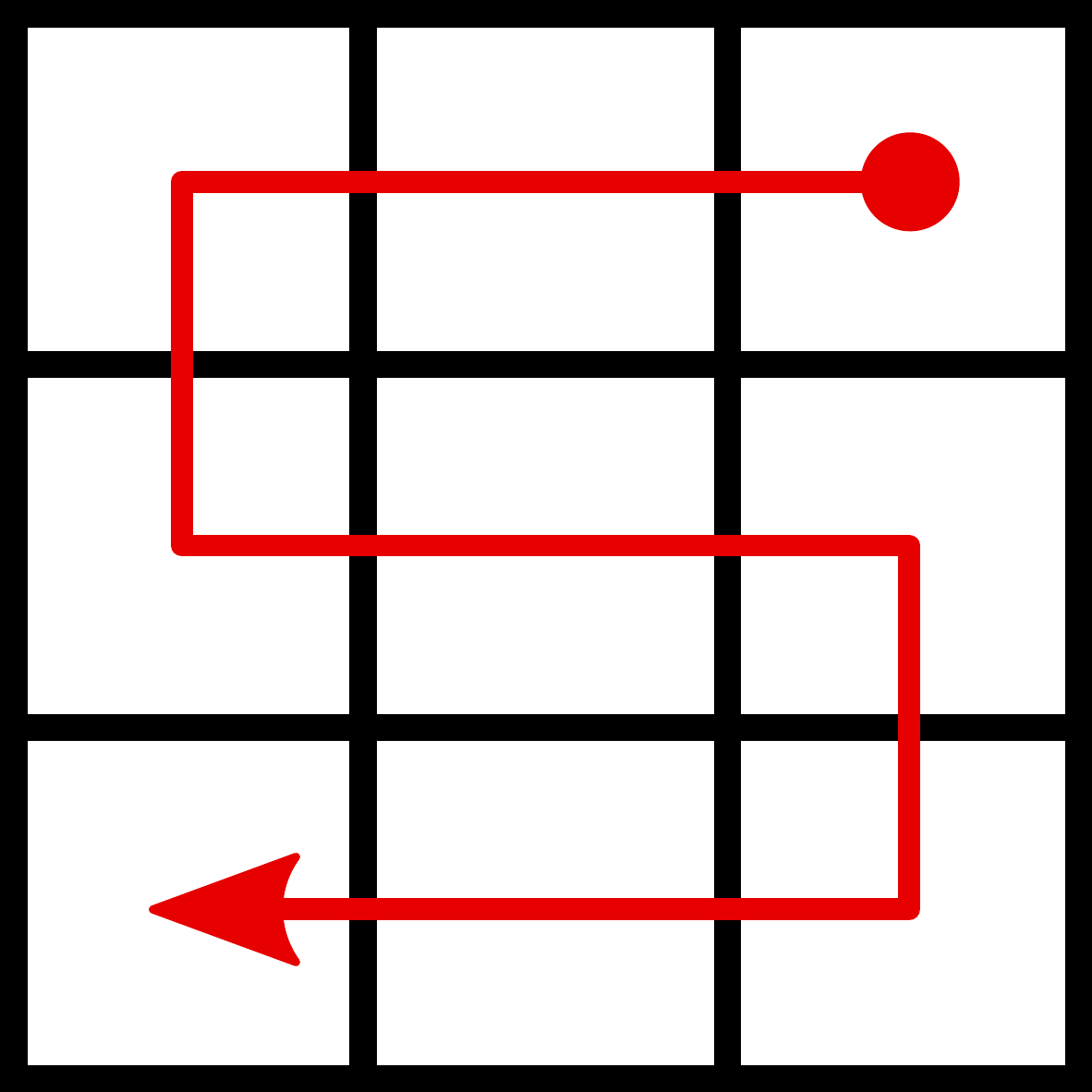}
    \label{fig:string3}}
    \subfigure[String 4.]{
    \includegraphics[width=0.16\linewidth]{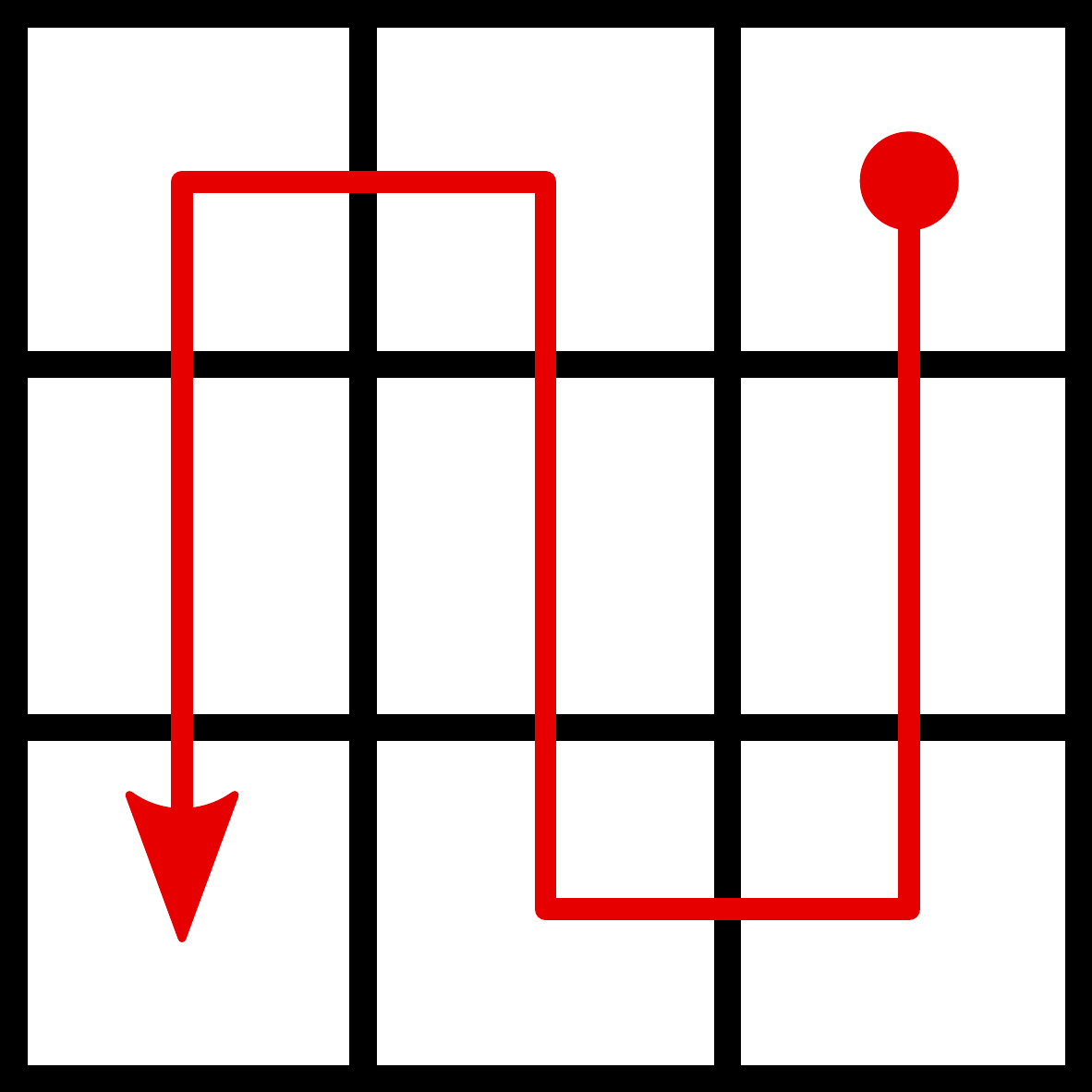}
    \label{fig:string4}}
    \subfigure[String 5.]{
    \includegraphics[width=0.16\linewidth]{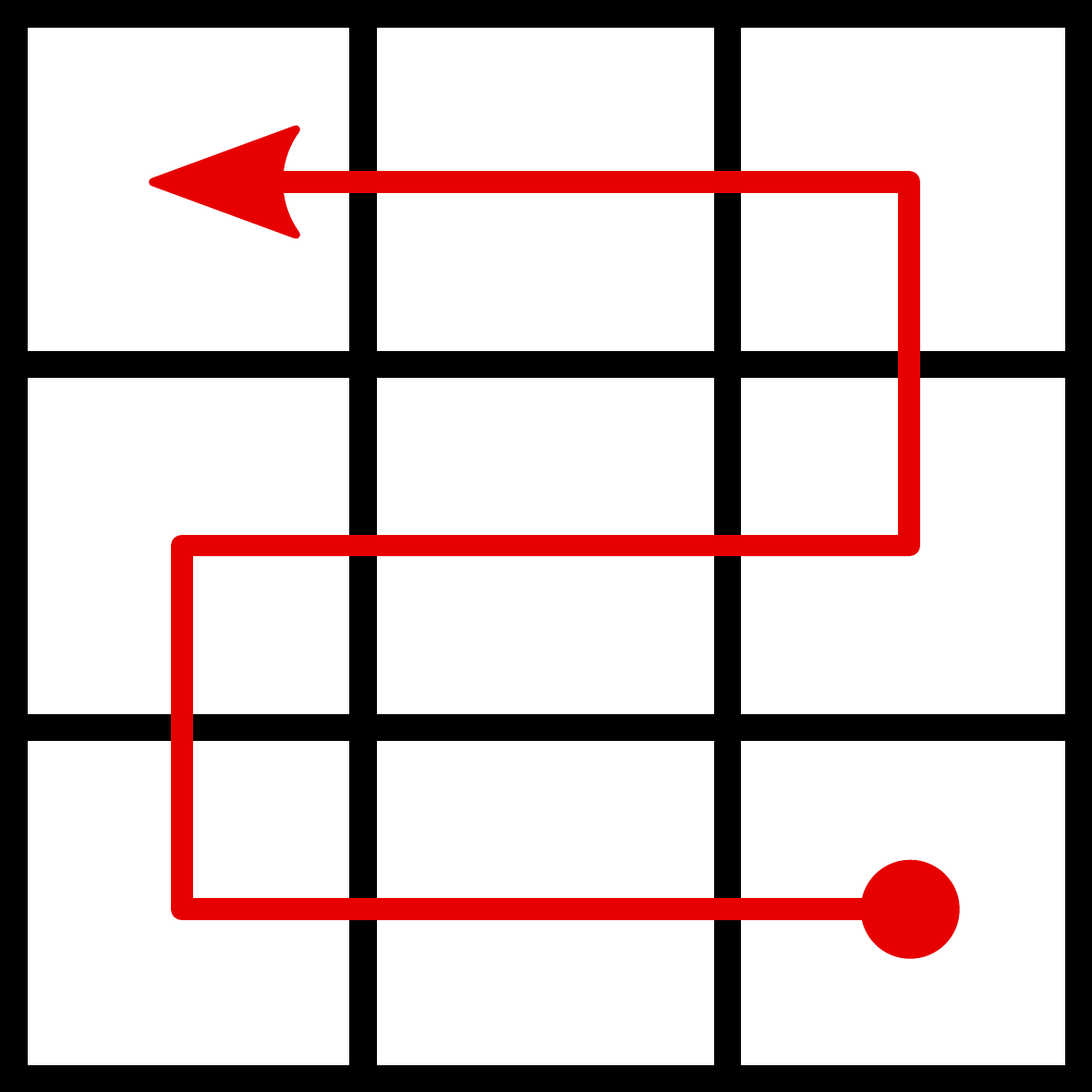}
    \label{fig:string5}}
    \subfigure[String 6.]{
    \includegraphics[width=0.16\linewidth]{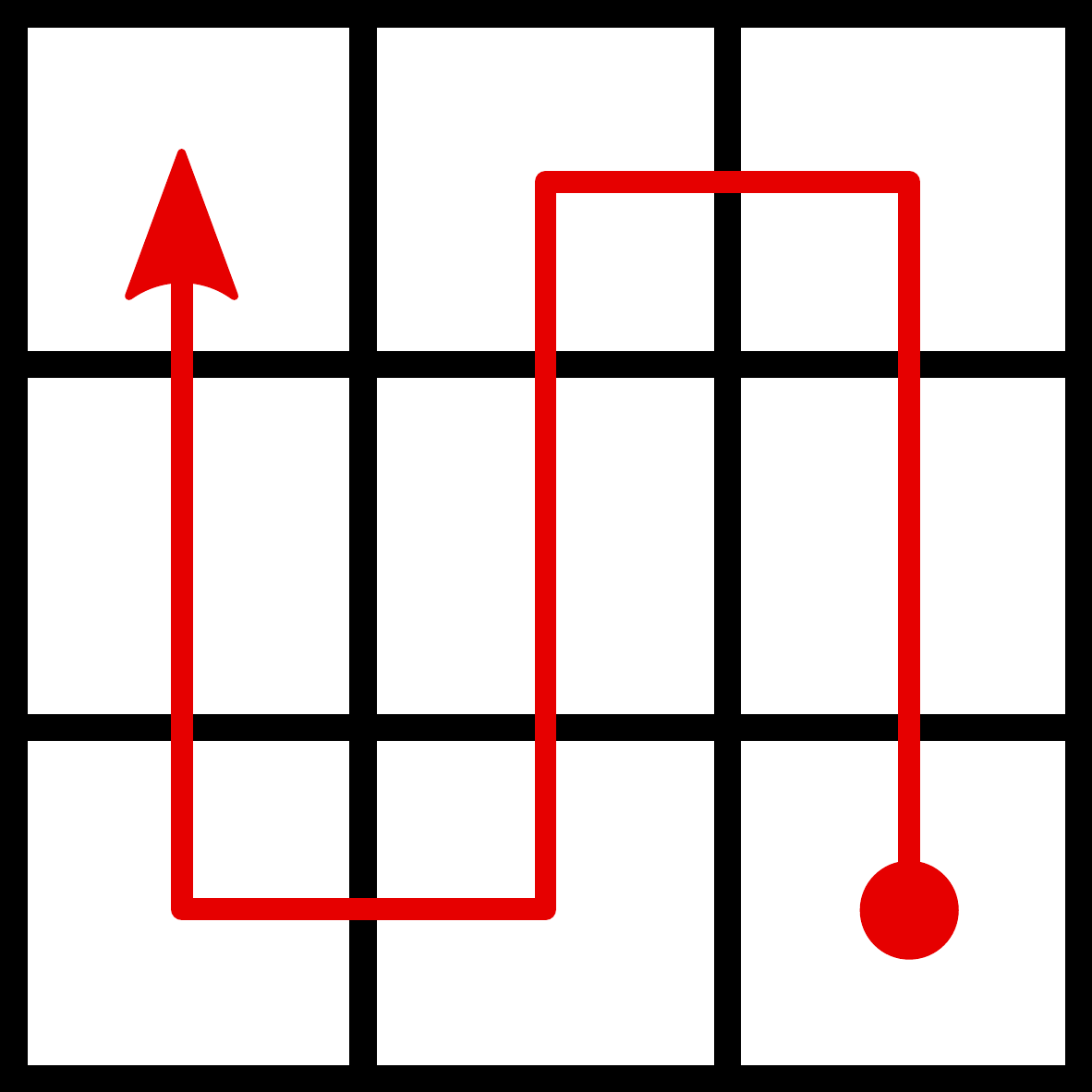}
    \label{fig:string6}}
    \subfigure[String 7.]{
    \includegraphics[width=0.16\linewidth]{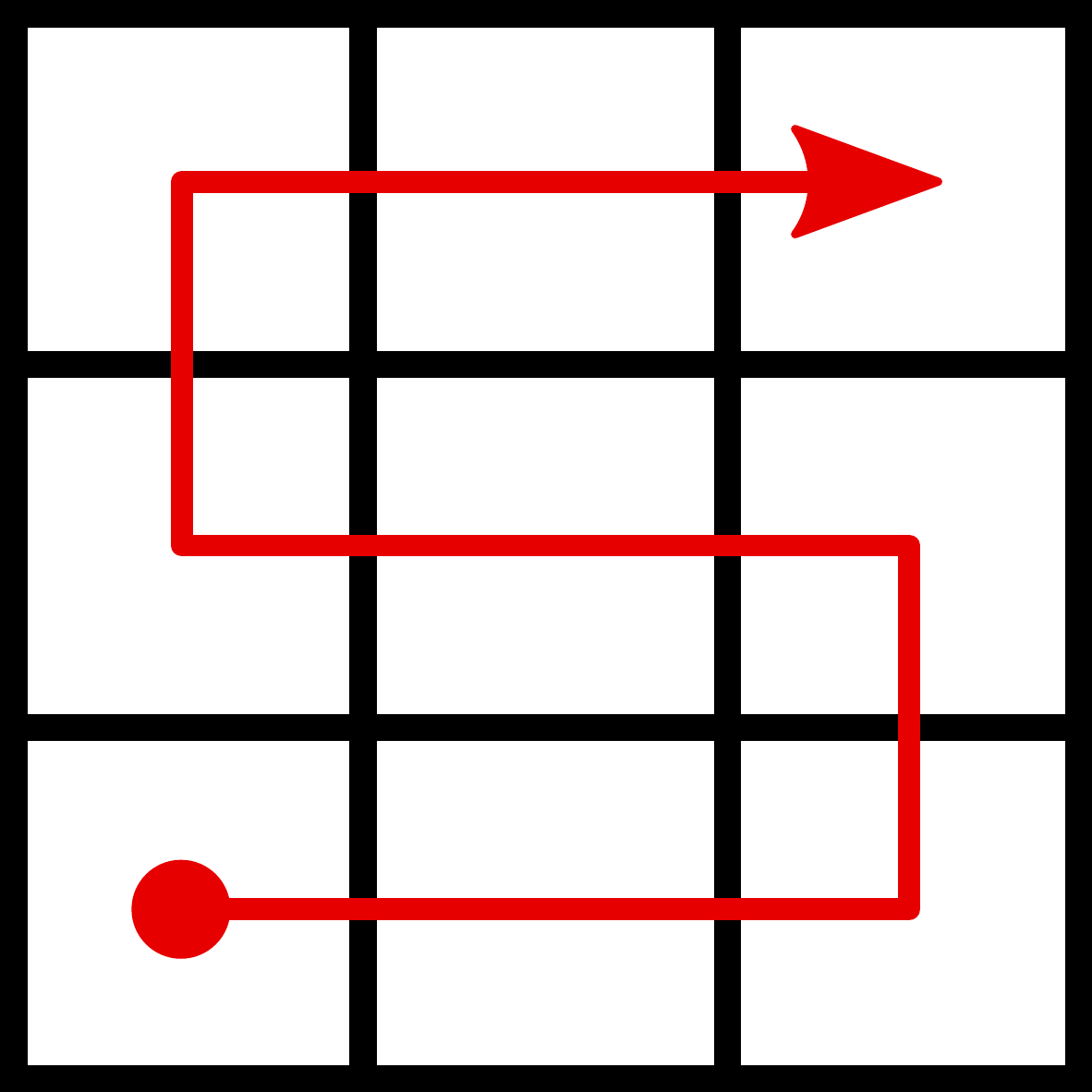}
    \label{fig:string7}}
    \subfigure[String 8.]{
    \includegraphics[width=0.16\linewidth]{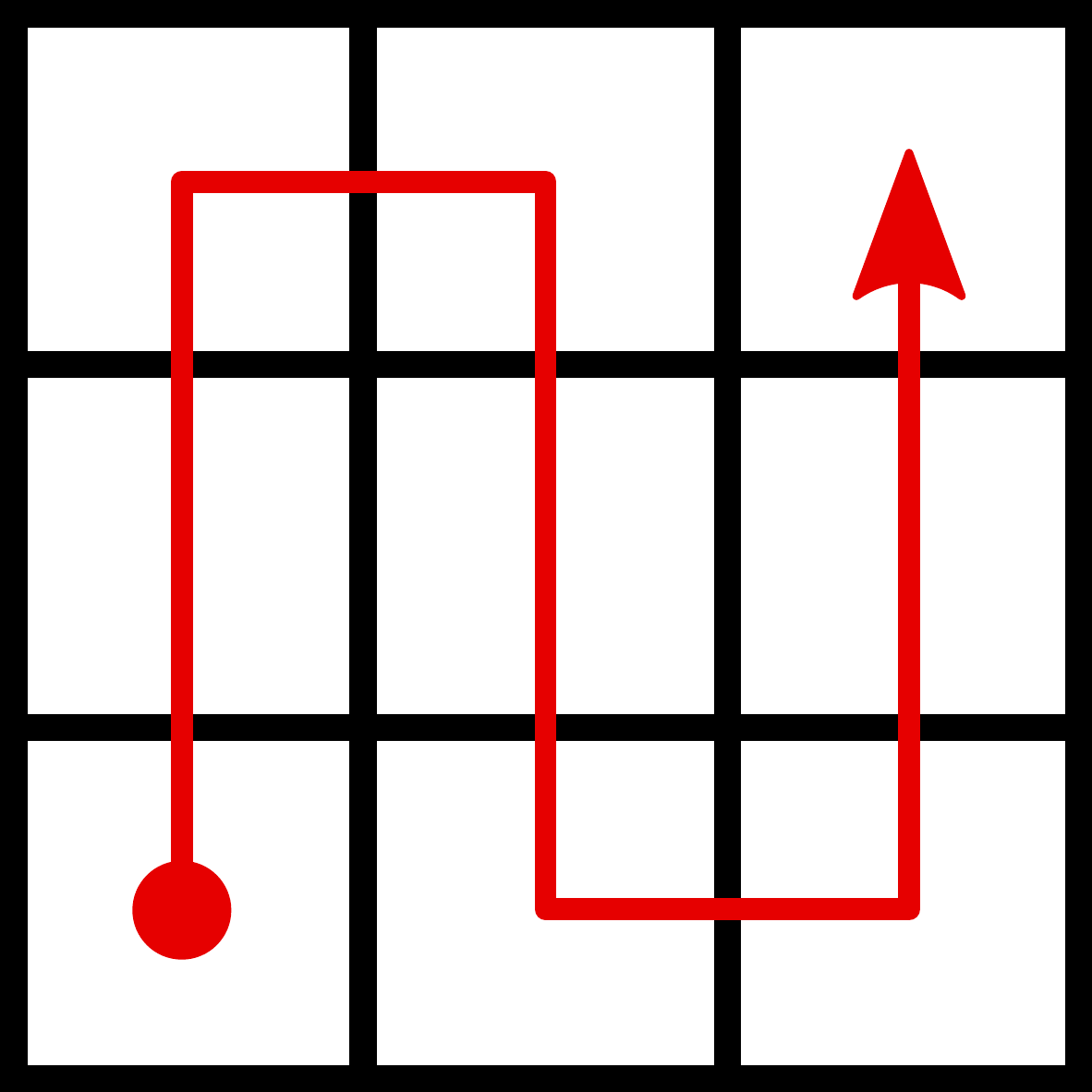}
    \label{fig:string8}}
    \caption{Strings used for mapping 1-D Transformer to 2-D quantum systems. String 0 is the default mapping (which we refer to as no strings). We always refer to the first (in order) $n$ strings (excluding string 0) when we say we used $n$ strings. }
    \label{fig:strings}
\end{figure}
Since our Transformer gives `ordered' measurement outcomes,  when we simulate two-dimensional systems we need to choose a linear ordering of our two-dimensional sites (i.e. a string of sites).  We consider two different single-string orderings (string 0 and string 1 from Fig.~\ref{fig:string0}).  
These strings explicitly break a symmetry of our system which then would need to be restored (to the degree to which the model has the variational freedom to do so) by the Transformer itself.  We can partially (or completely) restore this symmetry explicitly by choosing our ansatz to be a mixture of distributions defined over multiple different symmetry-related strings - i.e. $p_\theta(\bm{a}) = \sum_{\mathcal{S}} p_\theta(\bm{a}|\mathcal{S})p(\mathcal{S})$,
where $p(\mathcal{S}) = 1/N_{\text{string}}$ for a total number of $N_{\text{string}}$ strings; 
we call this refined ansatz a String state.
This linear combination of the Transformer probabilities can be interpreted as a mixture model \cite{hastie01statisticallearning} and bears some resemblance to string bond states \cite{PhysRevLett.100.040501}.
Restoring symmetries explicitly has proved useful in variational calculations of quantum  states \cite{mahajan2019symmetry,tahara2008variational,qiu2017projected,PhysRevResearch.2.023358,PhysRevB.100.125131}. 
Given a set of strings and a configuration $\bm{a}$ we can compute $p(\bm{a})$ explicitly. Sampling an $\bm{a}$ from $p_{\theta}(\bm{a})$ is also straightforward because of linearity and the fact that each term in our average is positive. To do so, we first sample an ordered $\{a_1,a_2,...a_{k-1}\}$ from the Transformer and then randomly choose a string to map these ordered values to get the final configuration. Here, we test a subset of strings $1$-$k$ for different $k$ (see Fig.~\ref{fig:string1}-\ref{fig:string8}).  

\textit{Optimization and Results.} Eq.~\ref{equ:lind-povm} gives a prescription for applying time evolution to the density matrix by time-evolving the POVM probability distribution.  To solve for the time-evolved distribution, we discretize time and use a second-order forward-backward trapezoid method \cite{iserles_2008}. We designed the following objective function 
\begin{align}
\begin{split}
    &\mathcal{C} = \frac{1}{N_s} \sum_{\bm{a} \sim p_{\theta(t+2\tau)}}^{N_s} \frac{1}{p_{\theta(t+2\tau)}(\bm{a})}\\
    &\bigg|\sum_{\bm{b}} \big[ p_{\theta(t+2\tau)}(\bm{b}) \left(\delta_{\bm{a}}^{\bm{b}} - \tau L_{\bm{a}}^{\bm{b}}\right)
    - p_{\theta(t)}(\bm{b}) \left(\delta_{\bm{a}}^{\bm{b}} + \tau L_{\bm{a}}^{\bm{b}}\right)\big]\bigg|,
    \label{eqn:dynamics}
\end{split}
\end{align}
where $N_s$ is the number of samples, $\delta_{\bm{a}}^{\bm{b}}$ is the Kronecker delta function, the sum over $\bm{a}$ is sampled stochastically from $p_{\theta(t+2\tau)}$, the sum over $\bm{b}$ can be evaluated effiencently as explained in Supplementary Material \cite{sup} Sec.~IX and the gradient of the objective function $\mathcal{C}$ with respect to the parameters in $p_{\theta(t+2\tau)}(\bm{b})$ is computed using PyTorch's \cite{paszke2019pytorch} automatic differentiation. To optimize the objective function we use Adam \cite{kingma2014adam}.
In the limit where $\mathcal{C}$ is zero, we get exact time evolution up to the discretization error induced by the trapezoid rule.  
More typically, it will be impossible for the Transformer to exactly represent the time-evolved state; instead by minimizing $\mathcal{C}$ the optimization continuously 
projects onto a nearby state in the manifold of distributions represented by our Transformer.  This can be viewed as a higher order generalization of IT-SWO \cite{kochkov2018variational} and the method in Ref.~\onlinecite{gutirrez2019real} but here applied instead to a probability distribution. The dominant source of error in performing our dynamics comes from the limited set of states that the Transformer can represent.  Additionally, it is possible that even within this manifold of states, one may not reach the optimal value if there are optimization issues such as local minima.  Over multiple time steps, errors will naturally accumulate due to the unitary dynamics of the system and be suppressed by the dissipative operators which should drive all dynamics to a fixed point.
\begin{figure}[h]
    \centering
    \includegraphics[width=\linewidth]{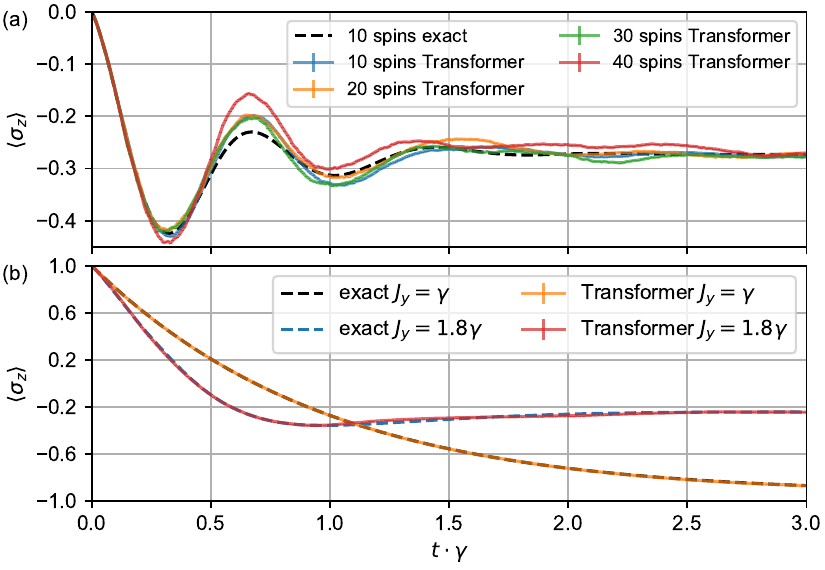}

    \caption{The expectation value $\langle \sigma_z\rangle$ as a function of time (a) for the 1-D Heisenberg model with $B = \gamma$, $J_x = 2\gamma$, $J_y = 0$, and $J_z = \gamma$ using a time step $\tau = 0.005 \gamma^{-1}$. The initial state is the product state  $\prod_{i=1}^N \ket{\leftarrow}$ ($\langle\sigma_y\rangle = -1$).  (b) for  the $3 \times 3$ Heisenberg model with $B = 0$, $J_x = 0.9\gamma$, $J_y = 1.0\gamma, 1.8\gamma$, and $J_z = \gamma$ using a time step $\tau = 0.008 \gamma^{-1}$. The initial state is the product state $\prod_{i=1}^N \ket{\uparrow}$ ($\langle\sigma_z\rangle = 1$).  Both models use periodic boundary conditions.  Exact curves are produced using QuTip \cite{JOHANSSON20131234,JOHANSSON20121760}.  The Transformer has one encoder layer and 32 hidden dimensions, and is trained using a forward-backward trapezoid method with a sample size $N_s=12000$.}
    \label{fig:10_low}
\end{figure}
We test this dynamic evolution on the 1-D and a 2-D Heisenberg model (see Fig.~\ref{fig:10_low}) using the tetrahedral POVM basis (see Supplementary Material \cite{sup} Sec.~II) where we find that the dynamics matches closely to the exact result. We capture both the qualitative behavior (i.e. the peaks and oscillation of the observables) as well as their quantitative values.   The values are especially accurate in both the limit of small and large time.  In our results, we have simulated one-dimensional chains up to $N=40$ and two-dimensional chains for $3 \times 3$ lattices.

One approach to finding the fixed point of the Liouvillian superoperator $\mathcal{L}$ is through a sufficiently long time-evolution (for an example see the large time limit of Fig.~\ref{fig:10_low}).  Interestingly, our approximate time evolution fluctuates around a fixed value of the observable, though it may not reach a true fixed point (i.e. $p_\theta(t+2\tau) = p_\theta(t)$) even in the limit of small $\tau$. (see Supplementary Material \cite{sup} Sec.~VII)

\begin{figure}[h]
    \centering
    \includegraphics[width=1\linewidth]{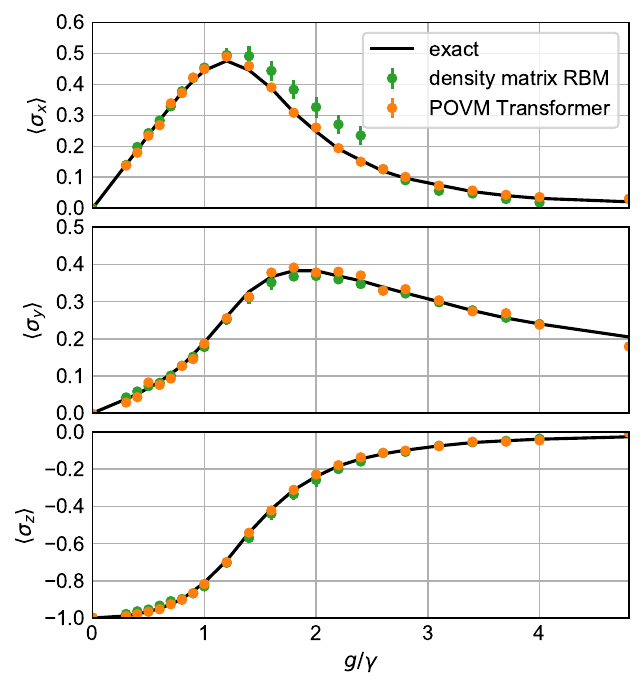}
    \caption{Variational steady-state solution for a 16-site TFIM chain with periodic boundary condition and $V = 2\gamma$ (orange dots). The initial state is the product state  $\prod_{i=1}^N \ket{\uparrow}$ ($\langle\sigma_z\rangle = 1$). The Transformer has one encoder layer and 32 hidden dimensions, and is trained using Adam \cite{kingma2014adam} in 500 iterations with $N_s=12000$. Green points are the fixed point solution representing the density matrix as an RBM;  both the exact curve (black line) and density matrix results are digitized from Ref.~\onlinecite{RBM_Vicentini}.}  
    \label{fig:16}
\end{figure}
\begin{figure}[h]
    \centering
    \includegraphics[width=1\linewidth]{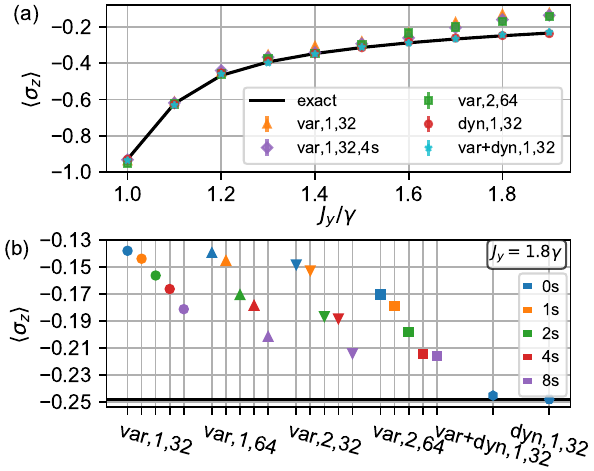}
    \caption{Steady-state solutions for $3 \times 3$ Heisenberg model with periodic boundary condition with $B = 0$, $J_x = 0.9\gamma$, and $J_z = \gamma$. The exact curves (black lines) are produced using QuTiP \cite{JOHANSSON20131234,JOHANSSON20121760}. (a) $\langle \sigma_z \rangle$ for different values of $J_y$ for POVM variational results (var), POVM dynamics (dyn) and POVM dynamics starting from the variational results (var+dyn).  The two integers in the legend label are the number of transformer layers and  hidden dimensions. (b) Steady-state solution $J_y=1.8\gamma$ comparing different variational ansatz. ``0s" and ``1s" use one string (String 0 and String 1); ``2s", ``4s", and ``8s" use Strings 1-2, 1-4, and 1-8 respectively (see Fig.~\ref{fig:string8}) All initial states are  $\prod_{i=1}^N \ket{\uparrow}$ ($\langle\sigma_z\rangle = 1$). The dynamics and variational plus dynamics approaches use the time step $\tau = 0.008 \gamma^{-1}$. The results of two transformer layers are computed exactly under all POVM frame elements.}
    \label{fig:3x3}
\end{figure}
Alternatively, we can search for the fixed point by direct minimization of the $L_1$-norm of $\dot{p_\theta}$ 
giving
\begin{equation}   
    \norm{\dot{p_\theta}}_1 =   \sum_{\bm{a}} \abs{\sum_{\bm{b}}p_\theta(\bm{b}) L_{\bm{a}}^{\bm{b}}}
      \approx  \frac{1}{N_s} \sum_{\bm{a} \sim p_\theta}^{N_s} \frac{\abs{\sum_{\bm{b}} p_\theta(\bm{b}) L_{\bm{a}}^{\bm{b}}}}{p_\theta(\bm{a})},
\label{eqn:L1norm}
\end{equation}
where the second line offers a stochastic approach to evaluate the $\norm{\dot{p_\theta}}_1$ by sampling $\bm{a}$ from $p_\theta({\bm{a}})$.  
The gradient in Eq.~\ref{eqn:L1norm} is taken with respect to the parameters in $p_{\theta}(\bm{b})$ using PyTorch's \cite{paszke2019pytorch} automatic differentiation. Notice that because the gradients of Eq.~\ref{eqn:dynamics} and Eq.~\ref{eqn:L1norm} (see Supplementary Material \cite{sup} Sec.~VII) are different (except in the limit where the manifold of states representable by the Transformer span the full space), they will converge to different answers. 

In Fig.~\ref{fig:16}, we consider the one-dimensional TFIM, with the 4-Pauli POVM basis (see Supplementary Material \cite{sup} Sec.~II), and compute the expectation value of all three Pauli matrices at various values of $\gamma$.  We find strong agreement with the exact method.  In addition, we find that this approach performs particularly well in the regime of $1<g/\gamma<2.5$ which have proven particularly challenging for the RBM method \cite{RBM_Vicentini}.  We can further improve the performance by averaging over multiple simulations (see Supplementary Material \cite{sup} Sec.~V). In Fig.~\ref{fig:3x3}, we consider optimizing a $3 \times 3$ Heisenberg model using Eq.~\ref{eqn:L1norm} with various different variational ansatz (here we use the tetrahedral POVM basis (see Supplementary Material \cite{sup} Sec.~II)).  In looking at the quality of $\langle \sigma_z\rangle$ we find that increasing the size of the Transformer both in depth and hidden dimension improves the result although this improvement is marginal until we reach two transformer layers and a hidden dimension of 64. Interestingly, we find that the use of strings has a significant effect on our results (see Fig.~\ref{fig:string8}). To begin with, the use of string 1 is marginally superior to string 0. We expect this is because string 1 better addresses local correlations.  More importantly, we find that  there is a significant improvement (for any Transformer) by including more symmetry related strings out to the maximum of eight strings we considered. In fact, eight strings with one hidden layer and a hidden dimension of 32  provides a similar accuracy to 1 string with 2 hidden layers and a hidden dimension of 64. Additionally, we compared the results obtained through time evolution at long time to the fixed point method and found that the steady-state approached by the time-evolved state provides significantly more accurate results. 
While the evaluation of the dynamics is computationally slower, we find that supplementing the fixed-point method with further dynamical evolution achieves the same steady-state solution as the dynamical approach at an overall reduced computational time.

\textit{Conclusion.} We have demonstrated an approach, whose run time complexity per iteration step is polynomial on the system size and the hidden dimensions, to simulate the real-time dynamics of open quantum systems via an exact probabilistic formulation. By parameterizing the quantum state using an autoregressive Transformer, we accurately track the dynamics and steady state in 1-D and 2-D transverse field Ising and Heisenberg models. For 2-D systems, we introduce String States which partially restore the symmetry of the Transformer. 

Our methods constitute an important step in the machine learning approach for quantum many-body dynamics simulation. It provides the first exact sampling method for neural networks in OQS, which is a crucial improvement over the standard Markov chain Monte Carlo techniques with RBM, as well as an efficient stochastic optimization method for high dimensional differential equations. Our approach is versatile and applicable to general quantum dynamics in various contexts, including closed systems quantum dynamics, finite temperature dynamics of the density matrix, as well as challenging fermionic transport problems \cite{PhysRevA.96.052110,yan2014} with interactions to the environment \cite{berkelbach2020}. Due to the probabilistic formulation as a quantum-classical mapping,  our work has applications beyond quantum mechanics and demonstrates how to efficiently solve high-dimensional probabilistic differential equations with autoregressive neural networks. Such probabilistic equations appear in a wide variety of classical contexts and our work represents an important step forward in the direction.

\textit{Acknowledgements.} Di Luo is grateful for the insightful discussion with Filippo Vicentini, and appreciates a lot the help from Filippo Vicentini, Alberto Biella and Cristiano Ciuti on providing the original data from their paper \cite{RBM_Vicentini}. Di Luo would also like to thank Mohamed Hibat-Allah for sharing his insights on the RNN wavefunction. Zhuo Chen is in debt to Qiwei Zhang for her contribution in digitizing Fig.~\ref{fig:16} for the exact result and drawing string figures in Fig.~\ref{fig:strings}. J.C. acknowledges support from Natural Sciences and Engineering Research Council of Canada (NSERC),  the Shared Hierarchical Academic Research Computing Network (SHARCNET), Compute Canada, Google Quantum Research Award, and the Canadian Institute for Advanced Research (CIFAR) AI chair program. BKC acknowledges support from the Department of Energy grant DOE desc0020165. This work utilized resources supported by the National Science Foundation’s Major Research Instrumentation program, grant \#1725729, as well as the University of Illinois at Urbana-Champaign”. Z.C. acknowledges support from the A.C. Anderson Summer Research Award.

\bibliography{reference}
\clearpage
\onecolumngrid
\renewcommand\thefigure{S\arabic{figure}}  
\renewcommand\thetable{S\arabic{table}}  
\renewcommand{\theequation}{S\arabic{equation}}
\renewcommand{\thepage}{P\arabic{page}} 
\setcounter{page}{1}
\setcounter{figure}{0}  
\setcounter{table}{0}
\setcounter{equation}{0}

\def\beq{\begin{equation}}
\def\eeq{\end{equation}}
\appendix
\section{\label{app:lindblad} \large{Supplementary Material for Autoregressive Transformer Neural Network for Simulating Open Quantum Systems via a Probabilistic Formulation}}

\section{I. Lindblad Equation in POVM Formalism}
Start from the Lindblad equation for density matrices
\begin{equation}
     \dot{\rho} = \mathcal{L}\rho
                = -i[{H}, \rho] + \sum_k\frac{\gamma_k}{2}\left(2\Gamma_k\rho \Gamma_k^\dagger - \{\rho, \Gamma_k^\dagger \Gamma_k\}\right),
    \label{equ:lind-rho}
\end{equation}
the frame and dual-frame satisfy
\begin{equation}
    p(\bm{a}) = \Tr(\rho M_{(\bm{a})}),
    \label{equ:rhoma}
\end{equation} 
and
\begin{equation}
    \rho = \sum_{\bm{b}} p(\bm{b})N^{(\bm{b})}.
    \label{equ:pbnb}
\end{equation}
Plugging Eq.~\ref{equ:pbnb} into Eq.~\ref{equ:lind-rho}, we have
\begin{equation}
    \sum_{\bm{b}} \dot{p}(\bm{b})N^{(\bm{b})} = \sum_{\bm{b}}p(\bm{b})\bigg[ -i [H, N^{(\bm{b})}]
    + \sum_{k}\frac{\gamma_k}{2}\left(2\Gamma_kN^{(\bm{b})} \Gamma_k^\dagger - \{N^{(\bm{b})}, \Gamma_k^\dagger \Gamma_k\}\right)\bigg].
    \label{equ:middle_step}
\end{equation}
Plugging Eq.~\ref{equ:pbnb} into Eq.~\ref{equ:rhoma}, we have
\begin{equation}
    p(\bm{a}) = \sum_{\bm{b}}p(\bm{b})\Tr(N^{(\bm{b})} M_{(\bm{a})}).
    \label{trmanb}
\end{equation}
Therefore, let's add $M_{(\bm{a})}$ and take trace on both sides of Eq.~\ref{equ:middle_step},
\begin{equation}
    \sum_{\bm{b}} \dot{p}(\bm{b})\Tr \left(N^{(\bm{b})}M_{(\bm{a})}\right)
    = \sum_{\bm{b}}p(\bm{b})\left[ -i \Tr\left([H, N^{(\bm{b})}]M_{(\bm{a})}\right)
    + \sum_{k}\frac{\gamma_k}{2}\Tr\left(2\Gamma_k N^{(\bm{b})} \Gamma_k^\dagger M_{(\bm{a})}
    - \{N^{(\bm{b})}, \Gamma_k^\dagger \Gamma_k\}M_{(\bm{a})}\right)\right].
\end{equation}
Replacing the left side with Eq.~\ref{trmanb} and rearranging the right side, we arrive at 
\begin{equation}
    \dot{p}(\bm{a}) = \sum_{\bm{b}}p(\bm{b})\left[-i\Tr\left({H}[{N}^{(\bm{b})}, {M}_{(\bm{a})}]\right) + \sum_k\frac{\gamma_k}{2}\Tr\left(2\Gamma_k{N}^{(\bm{b})}\Gamma_k^\dagger{M}_{(\bm{a})} - \Gamma_k^\dagger\Gamma_k\{{{N}^{(\bm{b})}}, {{M}_{(\bm{a})}}\}\right)\right] \equiv \sum_{\bm{b}}p(\bm{b}) L_{\bm{a}}^{\bm{b}}.
\end{equation}
Notice that the equation of motion is exact and mathematically equivalent to the standard density matrix Lindblad equation. Therefore, this equation preserves the positivity of the probability distributions as long as the initial probability distribution is positive and corresponds to a quantum state, which is the case as it is derived from a physical state. Our algorithm based on this equation also imposes positivity since the autoregressive neural network always parameterizes a positive probability distribution by construction.

\section{II. Tetrahedral and 4-Pauli POVM}
In the main paper, we used two POVMs, the tetrahedral POVM and the 4-Pauli POVM. The tetrahedral POVM forms a tetrahedral in the Bloch sphere. In particular, it takes the form of 
\begin{equation}
    M_{(a)} = \frac{1}{4} \left(\mathbbm{1} + \bm{v}_{(a)} \cdot \bm{\sigma}\right),
\end{equation}
where $\mathbbm{1}$ is the identity matrix, $\bm{\sigma}$ are the Pauli matrices, and $\bm{v}_{(a)}$ are four unit vectors which form a tetrahedral. In the main paper, we choose the four vectors to be
\begin{align}
    \bm{v}_{(1)} &= \left\langle 0, 0, 1\right\rangle, \\
    \bm{v}_{(2)} &= \left\langle \frac{2\sqrt{2}}{3}, 0, -\frac{1}{3}\right\rangle, \\
    \bm{v}_{(3)} &= \left\langle -\frac{\sqrt{2}}{3}, \frac{\sqrt{6}}{3}, -\frac{1}{3}\right\rangle, \\
    \bm{v}_{(4)} &= \left\langle -\frac{\sqrt{2}}{3}, -\frac{\sqrt{6}}{3}, -\frac{1}{3}\right\rangle.
\end{align}
The 4-Pauli POVM, on the other hand, takes the form of 
\begin{align}
    M_{(1)} &= \frac{1}{3}\ket{0}\bra{0},\\
    M_{(2)} &= \frac{1}{3}\ket{\raisebox{.4\height}{\scalebox{.6}{+}}}\bra{\raisebox{.4\height}{\scalebox{.6}{+}}},\\
    M_{(3)} &= \frac{1}{3}\ket{r}\bra{r},\\
    M_{(4)} &= \mathbbm{1} - M_{(1)} - M_{(2)} - M_{(3)},
\end{align}
where $\ket{0}$, $\ket{+}$, and $\ket{r}$ are the positive eigenstates of $\sigma_z$, $\sigma_x$, and $\sigma_y$ respectively. The multi-site POVM is constructed as the tensor product of single-site POVMs.

\section{III. Convergence of Loss}
\begin{figure}[h]
    \centering
    \includegraphics[width=0.55\linewidth]{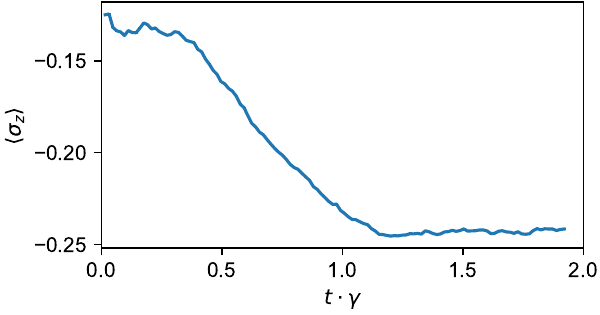}
    \caption{$\langle\sigma_z\rangle$ for variational plus dynamics result for $3 \times 3$ Heisenberg model with $B=0$, $J_x=0.9\gamma$, $J_y=1.8\gamma$, and $J_z=\gamma$ using one transformer layer, 32 hidden dimensions and no strings. The Transformer is trained using a forward-backward trapezoid method with 12000 samples and a time step $\tau = 0.008\gamma^{-1}$.}
    \label{fig:dynconverge}
\end{figure}
\begin{figure}[h]
    \centering
    \includegraphics[width=0.83\linewidth]{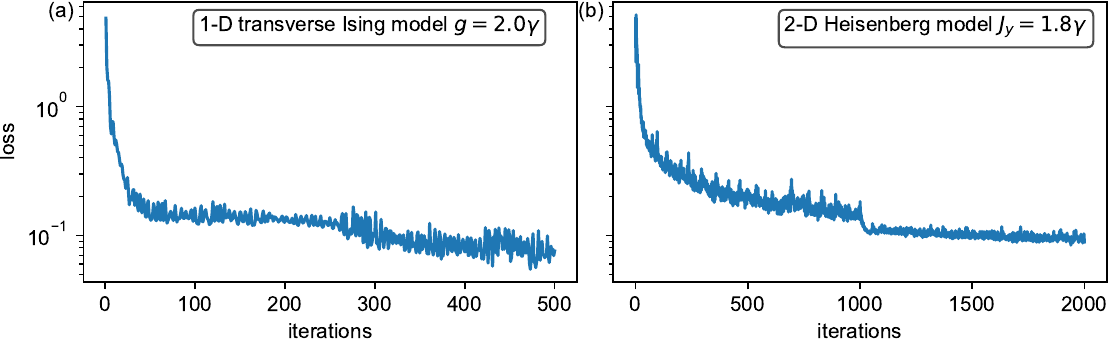}
    \caption{Examples of variational loss values. (a) loss values for 16 spins 1D TFIM with $V=g=2\gamma$ using one transformer layer and 32 hidden dimensions. (b) loss values for $3 \times 3$ Heisenberg model with $B=0$, $J_x=0.9\gamma$, $J_y=1.8\gamma$, and $J_z=\gamma$ using two layers, 64 hidden dimensions and no strings. The Transformer is trained using Adam \cite{kingma2014adam} with a sample size of 12000.}
    \label{fig:lossconverge}
\end{figure}

Here we show that the observable converges for dynamics process (Fig.~\ref{fig:dynconverge}) and the training loss converges for variational method (Fig.~\ref{fig:lossconverge}).

\section{IV. Results for Heisenberg Model with Larger Energy Scale}

\begin{figure}[h]
    \centering
    \includegraphics[width=0.83\linewidth]{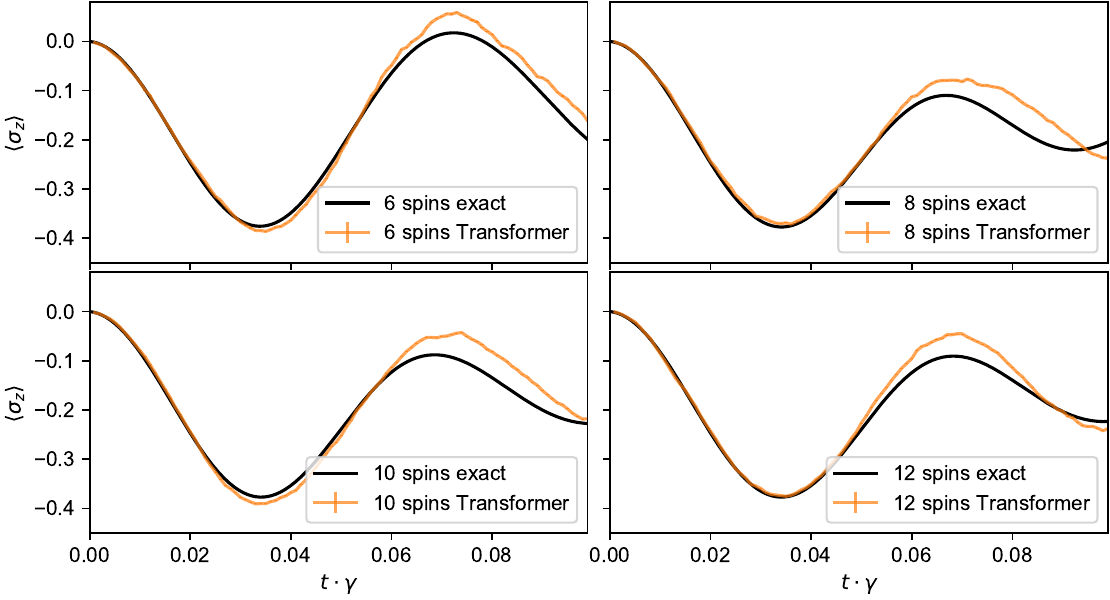}
    \caption{$\langle \sigma_z\rangle$ as a function of time computed with POVM dynamics. Various different system sizes of short time dynamics results for Heisenberg model in 1-D configuration with periodic boundary condition where $B = 10\gamma$, $J_x = 20\gamma$, $J_y = 0$, and $J_z = 10\gamma$. The initial state is a product state of $\prod_{i=1}^N \ket{\leftarrow}$ ($\langle\sigma_y\rangle = -1$). The exact curve is produced using QuTip \cite{JOHANSSON20131234,JOHANSSON20121760}. The Transformer is trained using a forward-backward trapezoid method with a sample size of 12000 and a time step of $0.0005\gamma^{-1}$. The neural network has one encoder layer and 32 hidden dimensions. }
    \label{fig:short}
\end{figure}
\begin{figure}[h]
    \centering
    \includegraphics[width=0.83\linewidth]{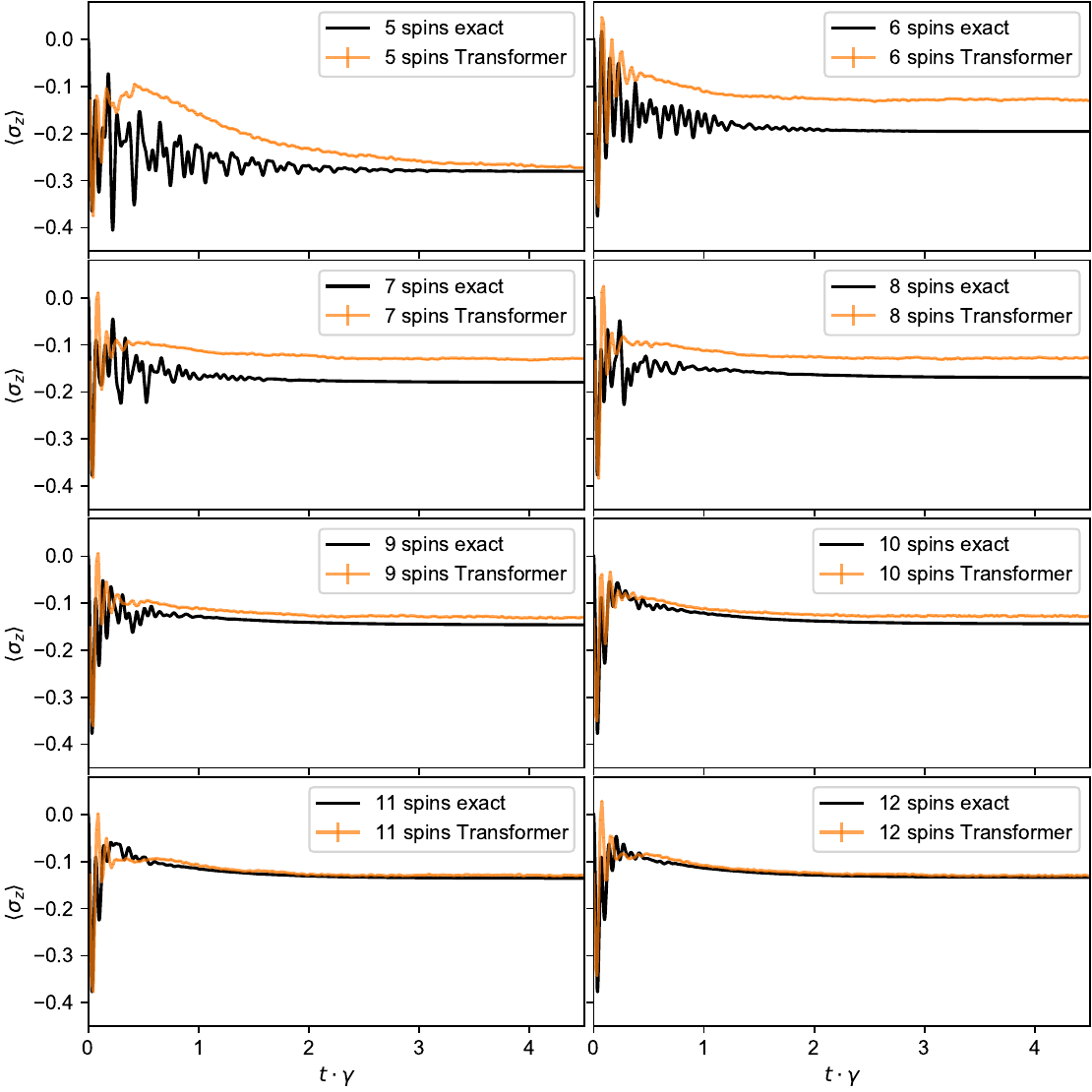}
    \caption{$\langle \sigma_z\rangle$ as a function of time computed with POVM dynamics. Various different system sizes of long time dynamics results for Heisenberg model in 1-D configuration with periodic boundary condition where $B = 10\gamma$, $J_x = 20\gamma$, $J_y = 0$, and $J_z = 10\gamma$. The initial state is a product state of $\prod_{i=1}^N \ket{\leftarrow}$ ($\langle\sigma_y\rangle = -1$). The exact curve is produced using QuTip \cite{JOHANSSON20131234,JOHANSSON20121760}. The neural network is trained using a forward-backward trapezoid method with a sample size of 12000 and a time step of $0.0075\gamma^{-1}$. The neural network has one encoder layer and 32 hidden dimensions. }
    \label{fig:long}
\end{figure}

The dynamics algorithm is also tested on a different Heisenberg model  where $B = 10\gamma$, $J_x = 20\gamma$, $J_y = 0$, and $J_z = 10\gamma$. This model has a higher energy compared with the model in the main paper. Since the dissipation operator is relatively small compared with the Hamiltonian, this model reveals closed system properties as well as open system properties. In Fig.~\ref{fig:short}, we show the short time dynamics for different number of spins of this model. It can be seen that for short time dynamics, the neural network predicts the observables to a great precision. In Fig.~\ref{fig:long}, we show the long time dynamics behavior for different number of spins. Even though the performances of different number of spins are different, it starts to converge to the right steady state as the system size goes larger.

\section{V. Improve Performance by Combining Probabilities}
\begin{figure}[h]
    \centering
    \includegraphics[width=0.83\linewidth]{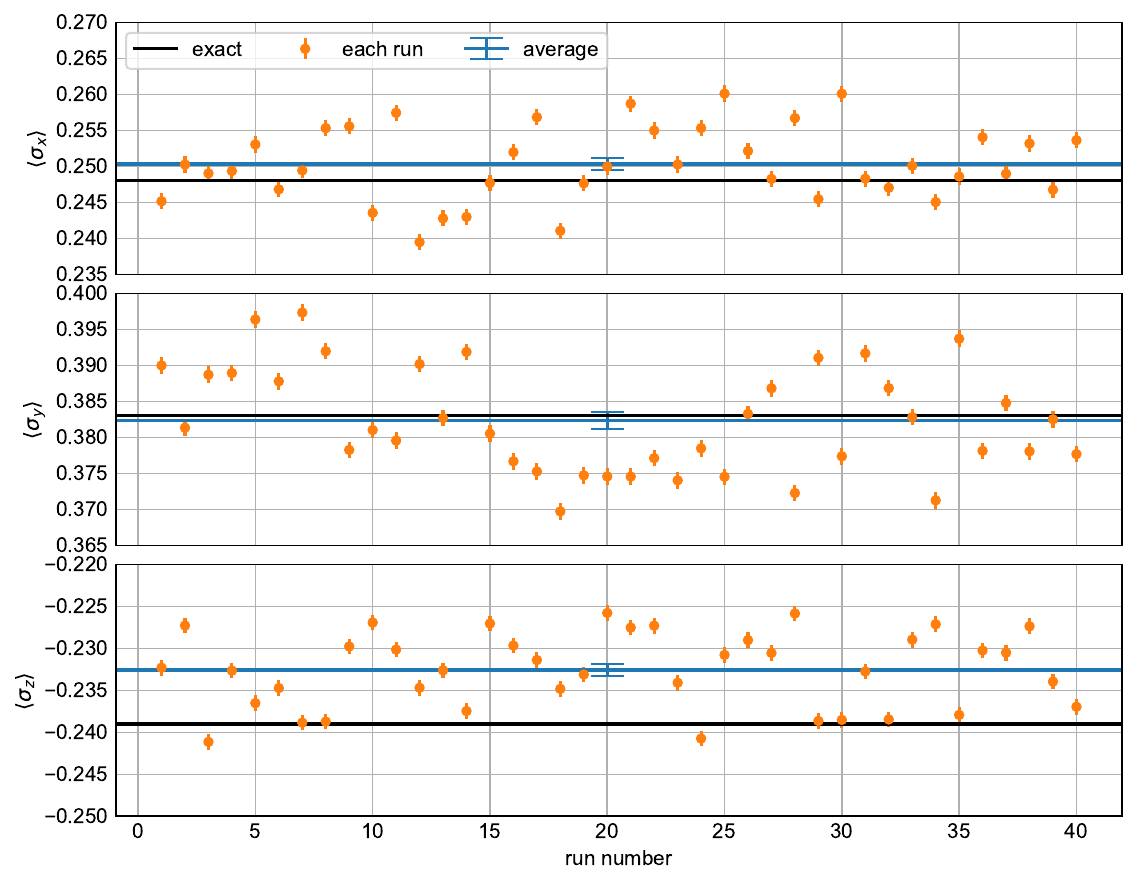}
    \caption{Data of multiple runs for variational steady state solution for one dimensional 16-site spin chain for the TFIM with periodic boundary condition where $V = g = 2\gamma$. The initial state is a product state of $\prod_{i=1}^N \ket{\uparrow}$ ($\langle\sigma_z\rangle = 1$). The neural network has one encoder layer and 32 hidden dimensions, and is trained using Adam \cite{kingma2014adam} in 500 iterations with a sample size of 12000. The exact curve (black line) is digitized from Ref.~\onlinecite{RBM_Vicentini}.}
    \label{fig:average}
\end{figure}
Because of the stochastic nature of the initialization and training process, each training could yield a slightly different result. In principle, we can average over multiple results to achieve better performance by defining the overall POVM probability
\begin{equation}
    p(\bm{a}) = \frac{1}{N}\sum_i p_i(\bm{a}).
\end{equation}
Then, the observable is computed as
\begin{equation}
    \langle O \rangle = \frac{1}{N}\sum_{i,\bm{b}}p_i(\bm{b})\Tr(O N^{\bm{b}}) = \frac{1}{N}\sum_i\langle O \rangle_i,
\end{equation}

which turns out to be the average of observables. In Fig.~\ref{fig:average}, we show the results for 1-D transverse Ising model for multiple training processes. It can be noted that for $\langle\sigma_x\rangle$ and $\langle\sigma_y\rangle$, the average result is indeed better.

\section{VI. Neural Network Initialization}
All the neural networks used in the main paper have the weights and biases (except in the last layer) initialized using PyTorch \cite{paszke2019pytorch} default linear layer normalization. 
All training process starts with a product state of either $\ket{\uparrow}$ or $\ket{\leftarrow}$ ($\langle\sigma_z\rangle = 1$ or $\langle\sigma_y\rangle = -1$ respectively, see figure captions for the exact initial state). To initialize the neural networks in such a product state, in the last fully connected layer, the weight is set to zero and the bias is set to $\log\left( \bra{\psi}M_{(a)}\ket{\psi}\right)$ where $M_{(a)}$ is the single spin POVM basis. Thus, after softmax, the output of the neural network would be the corresponding product state in POVM basis.

\section{VII. Explanation of the Dynamics and Variational Cost Functions in Detail}
In this section, we explain the dynamics and variational cost functions (Eq.~4 and Eq.~5) in the main paper in detail. We start with the dynamics cost function.
We would like to produce the probability distribution at $t+2\tau$ from the probability distribution at $t$ using the forward-backward trapezoid method \cite{iserles_2008} as
\begin{equation}
    p_{\theta(t+2\tau)}  - \tau L p_{\theta(t+2\tau)} = p_{\theta(t)}  + \tau L p_{\theta(t)}. \label{eq:fb}
\end{equation}
To make the notation compatible with the main paper, we could write Eq.~\ref{eq:fb} as
\begin{equation}
    \sum_{\bm{b}} p_{\theta(t+2\tau)}(\bm{b}) \left(\delta_{\bm{a}}^{\bm{b}} - \tau L_{\bm{a}}^{\bm{b}}\right) = \sum_{\bm{b}} p_{\theta(t)}(\bm{b}) \left(\delta_{\bm{a}}^{\bm{b}} + \tau L_{\bm{a}}^{\bm{b}}\right),
\end{equation}
where $\delta_{\bm{a}}^{\bm{b}}$ is the Kronecker delta function. Then, we could design the cost function as the $L_1$-distance between the left hand side and right hand side as
\begin{equation}
    \mathcal{C} = \sum_{\bm{a}}\bigg|\sum_{\bm{b}} \left[p_{\theta(t+2\tau)}(\bm{b}) \left(\delta_{\bm{a}}^{\bm{b}} - \tau L_{\bm{a}}^{\bm{b}}\right) - p_{\theta(t)}(\bm{b}) \left(\delta_{\bm{a}}^{\bm{b}} + \tau L_{\bm{a}}^{\bm{b}}\right)\right]\bigg|. \label{eq:exact_cost}
\end{equation}
Minimizing this cost function with respect to $p_{\theta(t+2\tau)}$ would be equivalent to solving the original equation if the cost function can be minimized to zero. However, since the neural network can only approximate the probability distribution, we are approximately solving the original equation. In addition, the probability distribution has a dimension that exponentially increases as the number of spins increases, so it is not feasible to minimize Eq.~\ref{eq:exact_cost} exactly. Therefore, we seek for a stochastic version of the cost function. This can be achieved by applying the trick of multiplying the cost function by $1 = p_{\theta(t+2\tau)}(\bm{a}) / p_{\theta(t+2\tau)}(\bm{a})$ as
\begin{equation}
    \mathcal{C} = \sum_{\bm{a}}p_{\theta(t+2\tau)}(\bm{a})\frac{1}{p_{\theta(t+2\tau)}(\bm{a})} \bigg|\sum_{\bm{b}} \left[p_{\theta(t+2\tau)}(\bm{b}) \left(\delta_{\bm{a}}^{\bm{b}} - \tau L_{\bm{a}}^{\bm{b}}\right) - p_{\theta(t)}(\bm{b}) \left(\delta_{\bm{a}}^{\bm{b}} + \tau L_{\bm{a}}^{\bm{b}}\right)\right]\bigg|. 
\end{equation}
Notice that we should only take gradient on $p_{\theta(t+2\tau)}(\bm{b})$ but not on $p_{\theta(t+2\tau)}(\bm{a})$. Then, we could turn the first $p_{\theta(t+2\tau)}(\bm{a})$ into sampling $\bm{a}\sim p_{\theta(t+2\tau)}$ and the resulting equation is

\begin{equation}
    \mathcal{C} = \frac{1}{N_s} \sum_{\bm{a} \sim p_{\theta(t+2\tau)}}^{N_s} \frac{1}{p_{\theta(t+2\tau)}(\bm{a})}
    \bigg|\sum_{\bm{b}} \big[ p_{\theta(t+2\tau)}(\bm{b}) \left(\delta_{\bm{a}}^{\bm{b}} - \tau L_{\bm{a}}^{\bm{b}}\right)
    - p_{\theta(t)}(\bm{b}) \left(\delta_{\bm{a}}^{\bm{b}} + \tau L_{\bm{a}}^{\bm{b}}\right)\big]\bigg|,
    \label{eqn:dynamics}
\end{equation}
which is exactly Eq.~4 in the main paper. Sampling $\bm{a}\sim p_{\theta(t+2\tau)}$ is efficient, as the autoregressive neural network is designed to sample from the probability distributions efficiently and exactly. (See Sec.~VIII for sampling details.) We should additionally notice that we don't need to sample over $\bm{b}$. Since the Hamiltonian and jump operators are local, $L_{\bm{a}}^{\bm{b}}$ is sparse. For a given $\bm{a}$, only a small number of (16 per two-body local Hamiltonian) $\bm{b}$'s are involved in the computation. Therefore, we could evaluate the sum over $\bm{b}$ exactly. Notice that autoregressive neural network allows exact inference so that the probability of each configuration can be exactly evaluated. The final cost function $C$ is then the expectation of the summand over $a$ sampled from $p_\theta(t+2\tau)$.

We can deal with the variational cost function similarly. Since we are searching for the steady state, we would like to solve for $p_\theta$ such that
\begin{equation}
    0 = \dot{p_\theta} = L p_\theta,
\end{equation}
or, in the notation of the main paper,
\begin{equation}
    0 = \dot{p_\theta}(\bm{a}) = \sum_{\bm{b}}p_\theta(\bm{b}) L_{\bm{a}}^{\bm{b}}. 
\end{equation}
We can just define the cost function as 
\begin{equation}
    \norm{\dot{p_\theta}}_1 = \sum_{\bm{a}}\abs{\sum_{\bm{b}} p_\theta(\bm{b}) L_{\bm{a}}^{\bm{b}}}.
\end{equation}
Similar to the dynamics cost function, this can be turned into a stochastic cost function using the same method, and the result is 
\begin{equation}
    \norm{\dot{p_\theta}}_1 = \frac{1}{N_s} \sum_{\bm{a} \sim p_\theta}^{N_s} \frac{\abs{\sum_{\bm{b}} p_\theta(\bm{b}) L_{\bm{a}}^{\bm{b}}}}{p_\theta(\bm{a})},
\end{equation}
the same as Eq.~5 in the main paper. Similar to the dynamics cost function, the gradient should be taken on $p_\theta (\bm{b})$ only.

Notice that these two cost functions serve two different purposes. The dynamics cost function is designed to train a neural network for each time step in an evolution, while the variational cost function seeks the steady state directly. In other words, the dynamics cost function produces many neural network states, where each of them represents the quantum state at a particular time, but the variational cost function only produces one neural network state, which is the final steady state. Since the evolution is dissipative, the neural network states generated from time evolution for large time should match the neural network state generated from the variational cost function. In practice, however, we noticed that time evolution generally produces better steady states. We believe the reason lies in the fact that the gradient of the two cost functions are different. (Please refer to the Sec.~IX for details.)

\section{VIII. Exact Sampling from Conditional Probability Distributions}
In this section, we explain how the probability distribution is sampled exactly. In the main paper, we explained that the Transformer neural network parameterizes the probability distribution over all spins as a product of conditional probabilities on each spin as
\begin{equation}
    p(\bm{a}) = p(a_1, a_2, a_3, \cdots) = \prod_k p_\theta(a_k| a_1, a_2, \cdots, a_{k-1}).
\end{equation}
The sampling procedure is as follows:
\begin{center}
    \begin{enumerate}
    \centering
    \item Sample $a_1'\sim p_\theta (a_1)$;
    \item Sample $a_2'\sim p_\theta (a_2|a_1')$;
    \item Sample $a_3'\sim p_\theta (a_3|a_1', a_2')$;
    \item[]$\vdots$
\end{enumerate}
\end{center}
Due to the autoregressive structure of the neural network, each sample can be drawn efficiently~\cite{transformer}. This procedure allows for sampling without Markov chain Monte Carlo (MCMC), so it does not need to ``warm up'' before generating usable samples. In addition it allows for an arbitrary number of samples to be sampled parallelly and independently, avoiding the correlation between samples in MCMC.

\section{IX. Efficient Evaluateion of Cost Functions}
Both the dynamics cost function
\begin{align}
\begin{split}
    \mathcal{C} = \frac{1}{N} \sum_{\bm{a} \sim p_{\theta(t+2\tau)}}^N \frac{1}{p_{\theta(t+2\tau)}(\bm{a})}
    \bigg|\sum_{\bm{b}} \big[ p_{\theta(t+2\tau)}(\bm{b}) \left(\delta_{\bm{a}}^{\bm{b}} - \tau L_{\bm{a}}^{\bm{b}}\right)
    - p_{\theta(t)}(\bm{b}) \left(\delta_{\bm{a}}^{\bm{b}} + \tau L_{\bm{a}}^{\bm{b}}\right)\big]\bigg|.
    \label{eqn:dynamicsappendix}
\end{split}
\end{align}
and the steady state cost function
\begin{equation}   
    \norm{\dot{p_\theta}}_1
      \approx  \frac{1}{N_s} \sum_{\bm{a} \sim p_\theta}^{N_s} \frac{\abs{\sum_{\bm{b}} p_\theta(\bm{b}) L_{\bm{a}}^{\bm{b}}}}{p_\theta(\bm{a})},
\label{eqn:L1norm}
\end{equation}
requires evaluating summation over $\bm{b}$. Since the operator ${{L_{\bm{a}}^{\bm{b}}}}$ is a sum of local operators, the cost functions need to sum only $\bm{b}$'s that is connected to the configuration $\bm{a}$'s through the local operators. For each $\bm{a}$, since the local operators only couple polynomial number of $\bm{b}$'s (specifically 16 in our case), the evaluation of the sum is efficient.

\section{X. Analysis of the Gradients of Dynamics and Variational Cost Functions}
The gradient of the dynamics cost function
\begin{align}
\begin{split}
    \mathcal{C} = \frac{1}{N} \sum_{\bm{a} \sim p_{\theta(t+2\tau)}}^N \frac{1}{p_{\theta(t+2\tau)}(\bm{a})}
    \bigg|\sum_{\bm{b}} \big[ p_{\theta(t+2\tau)}(\bm{b}) \left(\delta_{\bm{a}}^{\bm{b}} - \tau L_{\bm{a}}^{\bm{b}}\right)
    - p_{\theta(t)}(\bm{b}) \left(\delta_{\bm{a}}^{\bm{b}} + \tau L_{\bm{a}}^{\bm{b}}\right)\big]\bigg|.
    \label{eqn:dynamicsappendix}
\end{split}
\end{align}
is
\begin{align}
\begin{split}
    \frac{\partial \mathcal{C}}{\partial \theta} = \sum_{\bm{a}} \bigg[\sum_{\bm{b}}  &\frac{\partial p_{\theta(t+2\tau)}(\bm{b})}{\partial \theta} \left(\delta_{\bm{a}}^{\bm{b}} - \tau L_{\bm{a}}^{\bm{b}}\right) \bigg] \textrm{sign}
    \bigg\{\sum_{\bm{b}}\big[p_{\theta(t+2\tau)}(\bm{b}) \left(\delta_{\bm{a}}^{\bm{b}} - \tau L_{\bm{a}}^{\bm{b}}\right)
    - p_{\theta(t)}(\bm{b}) \left(\delta_{\bm{a}}^{\bm{b}} + \tau L_{\bm{a}}^{\bm{b}}\right)\big]\bigg\}
    \label{eqn:gradientdynamics}.
\end{split}
\end{align}
We find in our simulations that there is no numerical fixed point (i.e.  $p_{\theta(t)}=p_{\theta(t+2\tau)}$) for our Transformer dynamics.  This is plausible as  $\sum_{\bm{b}}  p_{\theta(t+2\tau)}(\bm{b}) L_a^b$ is not generically going to be zero except in the case when the Transformer reaches the exact steady state solution (which generically a moderate size transformer won't be able to represent).  This means that neither  Eq.~\ref{eqn:dynamicsappendix} nor Eq.~\ref{eqn:gradientdynamics} is going to be zero.

The gradient of the variational cost function 
\begin{equation}   
    \norm{\dot{p_\theta}}_1 =   \sum_{\bm{a}} \abs{\sum_{\bm{b}}p_\theta(\bm{b}) L_{\bm{a}}^{\bm{b}}}
      =  \frac{1}{N} \sum_{\bm{a} \sim p_\theta}^N \frac{\abs{\sum_{\bm{b}} p_\theta(\bm{b}) L_{\bm{a}}^{\bm{b}}}}{p_\theta(\bm{a})},
\label{eqn:L1normappendix}
\end{equation}
is

\begin{align}
    \frac{\partial \norm{\dot{p_\theta}}_1}{\partial \theta} = &  \sum_{\bm{a}}\left[\sum_{\bm{b}}\frac{\partial p_\theta(\bm{b})}{\partial \theta} L_{\bm{a}}^{\bm{b}}\right] \textrm{sign}\left[ \sum_{\bm{b}} p_\theta(\bm{b}) L_{\bm{a}}^{\bm{b}}\right].
    \label{eqn:gradientL1norm}
\end{align}
 
It is worth contrasting how these two approaches produce different results. It can be easily observed that the gradient is very different for these two approaches. In addition, the dynamics algorithm only locally matches the Transformer at two different time steps, while the variational algorithm globally searches for the steady state.

Since empirically the variational method is fast but not so accurate while the dynamics method is accurate but not fast, we might combine the gradient to take the advantages of both approaches. One can consider the interpolated dynamics cost function as follows
\begin{equation}
    \mathcal{C}_1 = \lambda \mathcal{C} + (1 - \lambda) \norm{\dot{p_\theta}}_1.
\end{equation}
During the dynamics process, one could slowly increase $\lambda$ from 0 to 1, switching from variational algorithm to dynamics algorithm. This cost function would produce inaccurate intermediate dynamics process, but should produce accurate steady state result. In the main paper, we have performed dynamics after the variational results, attaining accurate observables while reducing the training cost, which could be viewed as a special case of the above.

Even though the numerical fixed point may not be a local minimum of Eq.~\ref{eqn:dynamicsappendix} as discussed previously, it might still be useful to consider how gradient of the dynamics loss at the numerical fixed point looks like
\begin{align}
\begin{split}
    \frac{\partial \mathcal{C}}{\partial\theta} &= \sum_{\bm{a}} \left[\sum_{\bm{b}}  \frac{\partial p_\theta(\bm{b})}{\partial \theta} \left(\delta_{\bm{a}}^{\bm{b}} - \tau L_{\bm{a}}^{\bm{b}}\right)\right]\textrm{sign}\left[-\sum_{\bm{b}} p_\theta(\bm{b})L_{\bm{a}}^{\bm{b}}\right]\\
    &= \tau \frac{\partial \norm{\dot{p_\theta}}_1}{\partial \theta} - 
    \sum_{\bm{a}} \frac{\partial p_\theta({\bm{a}})}{\partial \theta}  \textrm{sign}\left[\sum_{\bm{b}} p_{\theta}({\bm{b}}) L_{\bm{a}}^{\bm{b}} \right].
    \label{eqn:gradientdynamicsend}
\end{split}
\end{align}
The first term is the same as the variational gradient, up to a scaling factor. Although a direct optimization using this gradient would not work since it only works at the numerical fixed point, one could be inspired by this gradient and formulate a new variational cost function as 
\begin{equation}
    \mathcal{C}_2 = \lambda \norm{\dot{p_\theta}}_1 - (1 - \lambda)\sum_{\bm{a}} p_\theta({\bm{a}})  \textrm{sign}\left[\sum_{\bm{b}} p_{\theta}({\bm{b}}) L_{\bm{a}}^{\bm{b}} \right].
\end{equation}
Then, one could choose different $\lambda$ to adjust the effect of the second term. In practice, we observed some improvements using this cost function, but the performances are unstable. It may be related to property that the sign function is sensitive to small changes.

\section{XI. Additional Benchmarks with Classical and Quantum Algorithms}
\begin{figure}[ht!]
    \centering
    \includegraphics[width=0.45\linewidth]{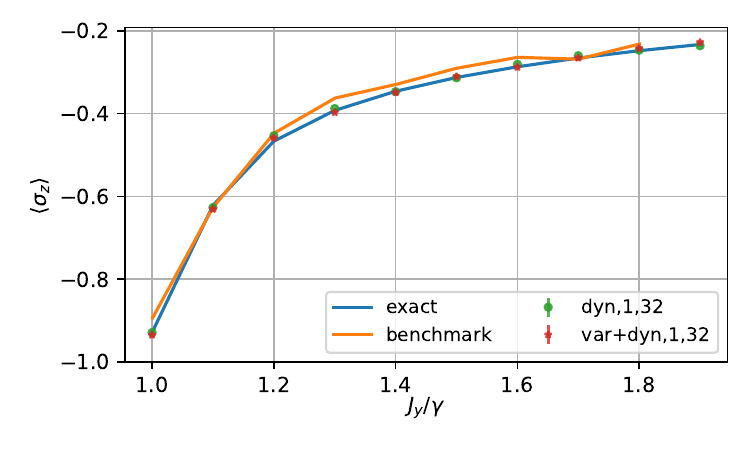}
    \caption{$3\times3$ Heisenberg model benchmarked with Ref.~\onlinecite{PhysRevLett.122.250501}. 
             This system is the same as in Fig.~4 in the main paper.
             The exact curve (blue) is generated using QuTiP~\cite{JOHANSSON20131234, JOHANSSON20121760}. 
             The benchmark curve (orange,  Ref.~\onlinecite{PhysRevLett.122.250501}), is based on an RBM.
             Our results (green and red) are the same as in the main paper. The two numbers in the legend specify the number of layers and hidden dimensions $n_d$.}
    \label{fig:rbm}
\end{figure}

\begin{figure}[ht!]
    \centering
    \includegraphics[width=\linewidth]{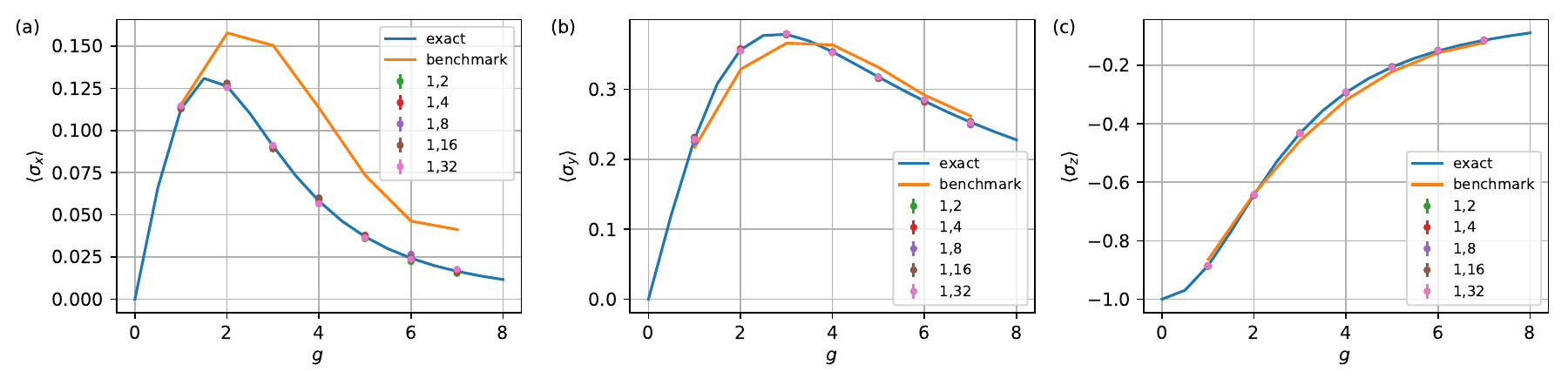}
    \caption{8 qubit transverse-field Ising model benchmarked with Ref.~\onlinecite{vqe_open}.
             The system Hamiltonian is the same as in Fig.~3 in the main paper but with open boundary condition, with $V=2$ and $g$ shown in the figure.
             The jump operators are slightly different from the main paper such that there are two different jump operators with $\Gamma^{(1)} = \sigma^{(-)}$,
             and $\Gamma^{(2)} = \sigma^{(z)}$. The corresponding dissipation rates are $\gamma^{(1)} = 4$ and $\gamma^{(2)} = 2$.
             Ref.~\onlinecite{vqe_open} uses a slightly different convention resulting in a difference in $g$ and $\gamma$ which we have verified by matching their curves in our convention.
             The exact curve (blue) is generated using QuTiP~\cite{JOHANSSON20131234, JOHANSSON20121760} 
             and is superimposed to the figure in Ref.~\onlinecite{vqe_open} to check for the correctness of the parameters. 
             The benchmark curve (orange) is from Ref.~\onlinecite{vqe_open}.
             The two numbers in the legend mean number of layers and hidden dimensions, respectively.
             We note that while the neural network is not designed to work for a hidden dimension $n_d$ less than 8, the results presented here are still significantly better than Ref.~\onlinecite{vqe_open} for $n_d<8$.}
    \label{fig:8qbit}
\end{figure}

\begin{figure}[ht!]
    \centering
    \includegraphics[width=0.8\linewidth]{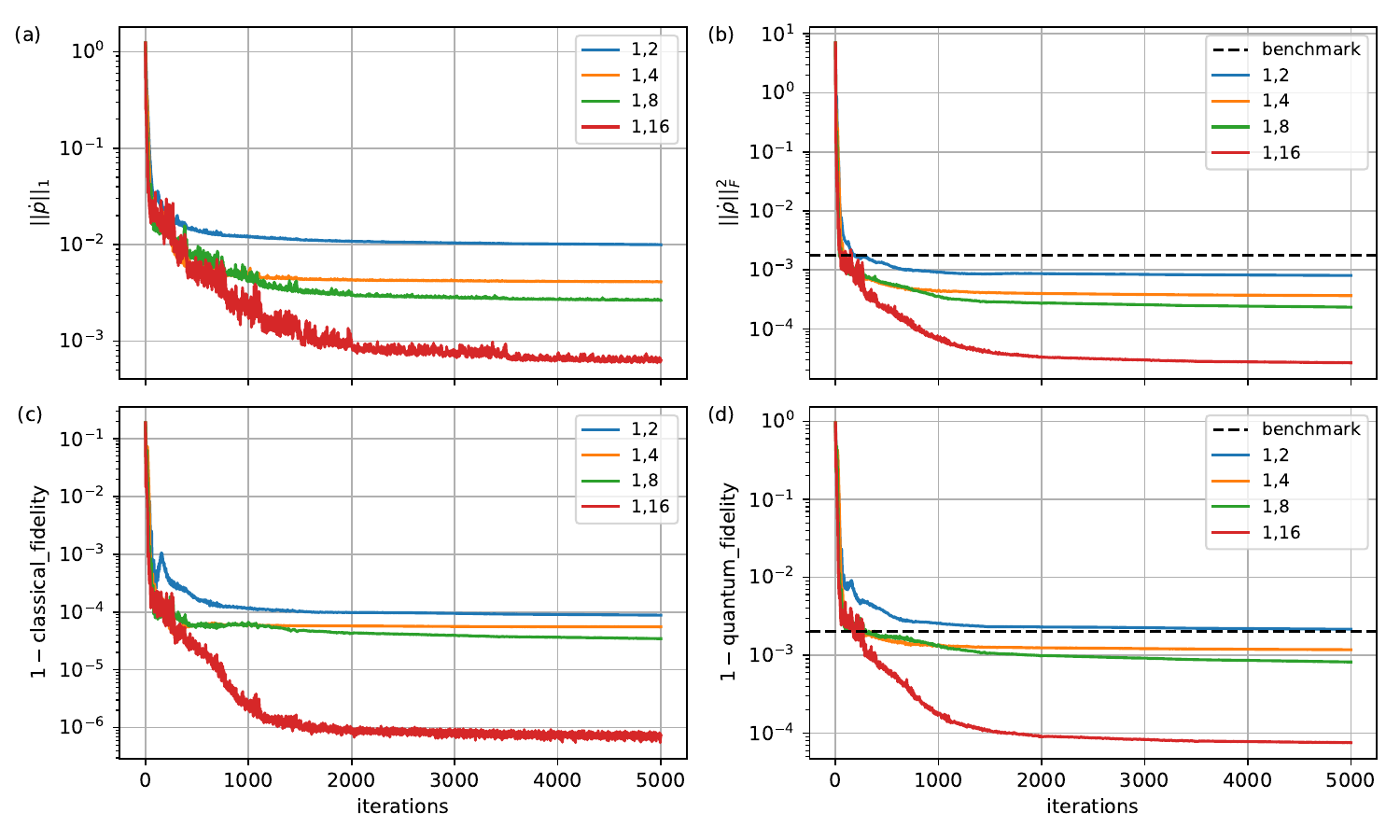}
    \caption{4 qubit transverse-field Ising model benchmarked with Ref.~\onlinecite{liu2021variational}.
             The system Hamiltonian and jump operators are the same as in Fig.~3 in the main paper, with $V=0.3$, $g=1$, and $\gamma=0.5$.
             The classical fidelity is defined as $(\sum_a\sqrt{p_\theta(a)p_{\text{exact}}(a)})^2$ 
             where $p_\theta$ is the neural network POVM probability distribution and $p_{\text{exact}}$ is the exact POVM probability distribution.
             The quantum fidelity is defined as $\Tr(\sqrt{\sqrt{\rho_\theta} \rho_\text{exact} \sqrt{\rho_\theta}})^2$,
             where $\rho_\theta$ is the density matrix converted from $p_\theta$ and $\rho_\text{exact}$ is the exact density matrix.
             The exact results are generated using exact linear solver. 
             The benchmark line (dashed black) is from Ref.~\onlinecite{liu2021variational}.
             The two numbers in the legend mean number of layers and hidden dimensions, respectively.
             We note that while the neural network is not designed to work for a hidden dimension  $n_d$ less than 8, the results presented here are still significantly better than Ref.~\onlinecite{liu2021variational} for $n_d<8$.}
    \label{fig:4qbit}
\end{figure}

In the main paper, we compared the results from Ref.~\onlinecite{RBM_Vicentini} 
and showed that we achieved better results. Here, we additionally benchmark with
Fig.~3 in Ref.~\onlinecite{PhysRevLett.122.250501} (results shown in Fig.~\ref{fig:rbm}), 
Fig.~9 in Ref.~\onlinecite{vqe_open} (results shown in Fig.~\ref{fig:8qbit}),
and Fig.~3 in Ref~\onlinecite{liu2021variational} (results shown in Fig.~\ref{fig:4qbit}).
Specifically, Ref.~\onlinecite{PhysRevLett.122.250501} is another stochastic machine learning algorithm with Restricted Boltzmann machine (RBM) in the standard density matrix formulation, while Ref.~\onlinecite{vqe_open} and Ref.~\onlinecite{liu2021variational} are recent variational quantum algorithms. Below, we show that our results are significantly better with respect to all the algorithms above. 

\end{document}